\documentclass[fleqn,usenatbib]{mnras}
\usepackage{amsmath,empheq,graphicx,txfonts,amssymb,natbib,epstopdf,grffile,hyperref,mathtools,lscape,array,color,afterpage}

\usepackage{tikz}
\usetikzlibrary{shapes,arrows}

\DeclareMathOperator{\diag}{diag}

\newcommand{\boldmu}{\boldsymbol{\mu}}
\newcommand{\boldSigma}{\boldsymbol{\Sigma}}
\newcommand{\boldx}{\boldsymbol{x}}

\def\apj{ApJ}
\def\apjl{ApJL}
\def\mnras{MNRAS}

\def\aap{A\&AP}

\def\apjs{APJS}

\def\araa{ARAA}
\def\ssr{SSR}

\title{Hierarchical Inference of the Relationship between Concentration and Mass in Galaxy Groups and Clusters}

\author[M. Lieu et al.]{\parbox{\textwidth}{Maggie Lieu,$^{1,2}$\thanks{E-mail: \href{mailto:maggie.lieu@esa.int}{maggie.lieu@esa.int}}
Will M. Farr,$^{1}$
 Michael Betancourt,$^{3,4}$
 Graham P. Smith,$^{1}$
  Mauro Sereno$^{5,6}$ and Ian G. McCarthy$^{7}$}\\
$^{1}$ School of Physics and Astronomy, University of Birmingham, Birmingham, B15 2TT, United Kingdom \\
$^{2}$ European Space Astronomy Centre (ESA/ESAC), Camino bajo del Castillo, E-28692 Villanueva de la Ca$\tilde{n}$ada, Madrid, Spain \\
$^{3}$ Applied Statistics Center, Columbia University, New York, NY 10027, USA  \\
$^{4}$ Department of Statistics, University of Warwick, Coventry CV4 7AL, United Kingdom \\
$^{5}$ INAF, Osservatorio Astronomico di Bologna, via Ranzani 1, 40127 Bologna, Italy \\
$^{6}$ Dipartimento di Fisica e Astronomia, Universit$\acute{a}$ di Bologna, viale Berti Pichat 6/2, I-40127 Bologna, Italy \\
$^{7}$ Astrophysics Research Institute, Liverpool John Moores University, Liverpool, L3 5RF, United Kingdom}

\pubyear{2016}

\begin{document}

\maketitle 
    
\begin{abstract}
  Mass is a fundamental property of galaxy groups and clusters. In theory weak gravitational lensing will enable an approximately unbiased measurement of mass, but parametric methods for extracting cluster masses from data require the additional knowledge of concentration. Measurements of both mass and concentration are limited by the degeneracy between the two parameters, particularly in low mass, high redshift systems where the signal-to-noise is low.
  In this paper we develop a hierarchical model of mass and concentration for mass inference we test our method on toy data and then apply it to a sample of galaxy groups and poor clusters down to masses of $\sim 10^{13}\, \rm M_\odot$. Our fit and model gives a relationship among masses, concentrations and redshift that allow prediction of these parameters from incomplete and noisy future measurements. Additionally the underlying population can be used to infer an observationally based concentration-mass relation.  Our method is equivalent to a quasi-stacking approach with the degree of stacking set by the data. We also demonstrate that mass and concentration derived from pure stacking can be offset from the population mean with differing values depending on the method of stacking.  \end{abstract}


\section{Introduction}
Galaxy groups and clusters are some of the largest structures in the observable Universe. They give insight to the growth and evolution of structure through the multi-wavelength study of their properties. Knowledge of the abundance and mass of these systems can be used in combination to probe cosmological parameters through the mass function \citep{Voit2005, Allen2011}. Although mass is not a direct observable, it can be estimated in a number of ways including hydrostatic mass from the X-ray emission of the hot intracluster medium and the dynamical mass from the velocity dispersions of galaxies. These estimators of mass rely on assumptions that may be biased from the true halo mass, for example X-ray masses could incur a bias of 10-30$\%$ \citep{Piffaretti2008, LeBrun2014} from the assumption of hydrostatic equilibrium. What's more, mass is generally observationally expensive.  

If gravity is the main contributor to the formation of clusters, then we would expect them to follow self-similarity \citep{Kaiser1986} and have simple power law relationships between mass and other observable properties known as mass proxies (temperature, luminosity etc.). These scaling relations, are a useful alternative to obtain mass measurements and are observationally cheaper. Nevertheless, scaling relations provide a less accurate estimate of mass and are influenced by the calibration cluster sample \citep{Sun2012, Giodini2013}.  

Weak lensing mass is a measure of the influence of the cluster gravitational potential on the light path of background galaxies \citep[see e.g.][for a review]{Hoekstra2013} and the arising galaxy shape distortion is known as shear. The effect is purely geometrical; it is sensitive only to line of sight structures and does not make as many assumptions as other methods, thus it provides a good estimator of the true halo mass. However lensing masses can suffer from the large scatter and noise. In particular, galaxy groups are $<10^{14}M_\odot$ making weak lensing measurements particularly challenging due to the low shear signal to noise ratio (SNR) and individual mass measurements in this context can be strongly biased \citep{Corless2007, Becker2011, Bahe2012}.

The NFW model \citep{Navarro1997} provides a reasonable description of the density profile of clusters, it is given by 
\begin{equation}\label{eqn: NFW}
\rho_{NFW}(r) = \frac{\rho_s}{(r/r_s)(1+r/r_s)^2},
\end{equation}
where $\rho_s$ is the central density and $r_s$ is a characteristic scale radius at which the slope of the log density profile is -2. The NFW model can be characterised by two parameters:  halo mass M$_\Delta$\footnote{M$_\Delta$ is the mass within which the mean density is $\Delta$ times the critical density at the cluster redshift} determines the normalisation and concentration $c_\Delta=r_\Delta/r_s$ determines the radial curvature of the profile. Whilst $M$ is both a physical quantity and a model parameter, $c$ is less well defined; $c$ is a parameter in the NFW profile but may not be equivalent in other density profiles \citep[e.g Einasto profile,][]{Klypin2014}. Concentration is difficult to constrain due its inherent covariance with mass \citep{Hoekstra2011,Auger2013, Sereno2015} and the degeneracy is particularly high for individual weak lensing measurements of high redshift, low mass systems.  Depending on the number of background galaxies, even massive clusters with reasonable shear SNR require the stacking of multiple clusters in order to constrain concentration \citep{Okabe2013, Umetsu2014}.The radial averaging when stacking helps to smooth out substructures; however it can be hard to decide which clusters to stack and how to stack them. What's more, stacking results in a loss of information. Therefore, it is more common to use a fixed concentration value \citep{Foex2012, Oguri2012, Applegate2014}, or a c--M scaling relation based on numerical simulations. \citep[e.g.][]{Bahe2012, Duffy2008, Dutton2014, Zhao2009}. The choice of c--M relation is again non-trivial, as dark-matter-only simulations tend to produce high normalisation relations compared to those that include baryonic physics and feedback \citep[e.g.][]{Duffy2010, Velliscig2014}. It is also sensitive to $\sigma_8$ and $\Omega_M$, where \cite{Duffy2008}'s c--M relation (which assumes a WMAP5 cosmology) has 20$\%$ lower concentrations than \cite{Dutton2014}'s relation (which assumes the Planck 2013 cosmology). These issues will affect  both mass and concentration due to parameter degeneracies.

Accurate mass measurements are important for cluster cosmology, however traditionally, methods to obtain cosmological constraints from the data are divided into separate analyses and work from the bottom-up. For example, observations are made and are processed into data catalogues, the catalogues are used to obtain individual masses of some clusters where the data quality is adequate to do so, a scaling relation fit is obtained for some mass proxy to allow further mass estimates of clusters where the data quality for mass is poor, and finally the cosmology can be obtained by fitting a mass function. Not only is this inefficient, it is also sub-optimal due to the loss of information, introduction of biases and the difficulty in consistent propagation of uncertainties at each step. 

Here we instead consider a Bayesian inference model that embeds the global problem into a forward modelling approach and subsequently avoids these many issues. Hierarchical modelling is a unified statistical analysis of the source population and individual systems. The prior distribution on the individual cluster parameters can seen as a common population distribution and the data can collectively be used to infer aspects of the population distribution that is otherwise not observed. In traditional non-hierarchal methods, introducing too few model parameters produces inaccurate fits to large data sets and too many parameters runs the risk of overfitting the data. By treating the problem as a hierarchical model \citep[see e.g.][]{Schneider2015, Alsing2015, Sereno2016} we have enough parameters to fit the data well when possible; the population distribution accounts for a full statistical dependence of all parameters  when not otherwise constrained by data. This ``quasi-stacking" approach enables improved estimates on weakly constrained parameters such as concentration and masses of low SNR clusters by incorporating information from the population in a principled way.

In this paper we propose a method to exploit the underlying cluster population properties in order to improve constraints on weak lensing masses of individual groups and poor clusters. The data are fit with the assumption that the parameters originate from the same underlying population. In the case of mass and concentration, the distribution of the population mass and concentration is a prior on the corresponding individual cluster parameters. This method is therefore fully self-consistent with the data and makes it possible to constrain concentration of each cluster without the need of full stacking. It works well even with low signal to noise data which will be important for future weak lensing surveys where the observations may be shallow such as DES\footnote{http://www.darkenergysurvey.org} and KIDS\footnote{http://kids.strw.leidenuniv.nl}.

This paper is structured as follows: in \autoref{sec:Method} we describe in detail the hierarchical model and outline how it can be used for parameter prediction. This is tested on toy data and simulations in \autoref{sec: toydata} and applied to a typical shallow data sample in \autoref{sec:Results}. Finally we conclude in \autoref{sec:Conclusion}. Throughout, the WMAP9 \citep{Hinshaw2013} cosmology of $H_0 = \rm 70\, km\, s^{-1}\, Mpc^{-1}$, $\Omega_{\rm M} = 0.28$ and $\Omega_\Lambda = 0.72$ is assumed. All statistical errors are reported to 68$\%$ credibility and all mass values are reported in units $h_{70}^{-1}\ \rm M_\odot$, unless otherwise stated.

\section{Method}\label{sec:Method} 
Our model assumes each cluster can be described by $n$ parameters. We assume that the distribution of the parameters for a population of clusters is described by a multivariate gaussian with a global mean $n$-vector $\boldmu$ and a $n\times n$ covariance matrix $\boldSigma$ that describes the intrinsic scatter of each property and the covariances between them. For now, we focus on the cluster mass M$_{200}$, concentration c$_{200}$ and redshift z. 
Therefore n=3,
\begin{equation}
	\boldmu =  \rm \left( \begin{array}{c}\overline{\ln(M_{200})} \\ \overline{\ln(c_{200}}) \\ \overline{\ln(1+z)} \end{array} \right), \nonumber
\end{equation}
and
\begin{equation}
	\boldSigma =	 \left( \begin{array}{ccc} \Sigma_{11} & \Sigma_{12}& \Sigma_{13}\\ 
					\Sigma_{21} & \Sigma_{22}& \Sigma_{23}\\ 
					\Sigma_{31} & \Sigma_{32}& \Sigma_{33} \end{array} \right) = 
	 \left( \begin{array}{ccc}\rm \sigma_{1}^{2} &\rm \rho_{12} \sigma_{1}\sigma_{2} &\rm \rho_{13} \sigma_{1}\sigma_{3}\\ 
					\rm \rho_{12} \sigma_{1}\sigma_{2} &\rm \sigma_{2}^{2} & \rho_{23} \sigma_{2}\sigma_{3}  \\ 
					\rm \rho_{13} \sigma_{1}\sigma_{3} &\rm \rho_{23} \sigma_{2}\sigma_{3}  & \sigma_{3}^{2} \end{array} \right). \nonumber
\end{equation}
Here the subscripts 1, 2, 3 on $\sigma$ represent $\ln M_{200}$, $\ln c_{200}$ and $\ln(1+z)$ respectively and $\rho$ is the correlation coefficient.

The true distribution for the population mass should be the cluster mass function which describes the number density of clusters of a given mass and redshift \citep[e.g.][]{Tinker2008}. Massive clusters form from rare, dense peaks in the initial mass-density fluctuations of early Universe so are less abundant than poor clusters and low mass groups that form from smaller more common fluctuations. However the least massive systems are also the least luminous and are therefore less likely to be detected than more luminous massive clusters. This selection function causes a decrease in the number of clusters observed at low mass due to survey sensitivity limits. Although in detail, the cluster selection function will not be log-normal, here we justify the use of a log-normal distribution as an approximation to the cluster mass function and selection function \citep[see also][]{Sereno2015}. This is also motivated for conjugacy since simulations of the cluster concentration mass distribution shows log-normal scatter \citep{Jing2000, Bullock2001, Duffy2008, DeBoni2013} and the results of \cite{Lieu2016} show redshift and mass distributions that are close to gaussian.

\subsection{Hyperparameters}
It is common to call the parameters that describe the population ($\mu$ and $\Sigma$) hyper-parameters, and the priors on them, hyper-priors. The covariance matrix $\Sigma$ is a difficult parameter to sample since by definition it must be both symmetric and positive definite. Therefore for its prior we take the \cite{Stan-manual2015} recommended approach, which decomposes $\Sigma$ into a correlation matrix $\boldsymbol{\Omega}$ and a scale vector $\boldsymbol{\tau}$  \citep{Barnard2000}: 
\begin{equation}
	\boldSigma = \rm diag(\boldsymbol{\tau}) \boldsymbol{\Omega} diag(\boldsymbol{\tau})
\end{equation}
$\boldsymbol{\tau}$ is a vector of the standard deviations of the hyper parameter $\mu$ which describe the population mean.
The prior on $\boldsymbol{\tau}$ is taken to be a Gamma distribution with shape $\alpha_\tau=2$ and rate $\beta_\tau=3$. 
\begin{equation}
	\mathrm{Pr}(\tau|\alpha_\tau, \beta_\tau) = \frac{\beta_\tau^{\alpha_\tau}}{\Gamma(\alpha_\tau, 1)}\tau^{\alpha_\tau-1}\exp(-\beta_\tau \tau),
\end{equation}
This prior is chosen so as to prevent divergences (see \autoref{sec:Model fitting}) in the sampling whilst allowing large values of variance. An LKJ distribution prior \citep{Lewandowski2009} is used on the correlation,
\begin{equation}
	\mathrm{Pr}(\boldsymbol{\Omega}|\nu) \propto \mathrm{det}(\boldsymbol{\Omega})^{\nu-1}.
\end{equation}
where the shape parameter $\nu>$ 0. This distribution converges towards the identity matrix as $\nu$ increases, allowing the control of the correlation strength between the multiple parameters and consequently the variance and covariance of parameters in the population. A flat prior can be imposed by setting $\nu$=1 and for $0<\nu<1$ the density has a trough at the identity matrix. However to optimise our code we decompose the correlation matrix $\boldsymbol{\Omega}$ into its Cholesky factor $\boldsymbol{L}_\Omega$ and its transpose $\boldsymbol{L}_\Omega^\intercal$,
\begin{align}
&	\boldsymbol{\Omega} = \boldsymbol{L}_\Omega^{} \boldsymbol{L}_\Omega^\intercal \\
& 	\mathrm{Pr}(\boldsymbol{\Omega}|\nu) = \prod_{k=2}^K L_{kk}^{K-k+2\nu-2},
 \end{align}
 and implement on $\boldsymbol{L}_\Omega$ a LKJ prior parameterised in terms of the Cholesky decomposition setting $\nu$=10, i.e. weakly preferring identity. For the global mean vector we use a weakly informative prior
 \begin{equation}
	\mathrm{Pr}(\boldmu|\boldmu_0, \boldSigma_0) = \frac{1}{\sqrt{2\pi\boldSigma_0}} \exp\left[ - \frac{(\boldmu-\boldmu_0)^2}{2\boldSigma_0} \right],
\end{equation}
where $\boldmu_0$=(32,1,0.3) and $\boldSigma_0=(1,1,1)$. 
The priors and hyper-priors chosen in our model are consistent with the knowledge of these systems since we expect masses to lie between $10^{13-16}\ \rm M_\odot$ and concentration between 0-10. Using the prior information helps to regularise the inference and avoids numerical divergences since the projected NFW profile is numerically unstable (in particular at the Einstein radius). We test the sensitivity of our results to these choices of hyperpriors in \autoref{sec:testpriors}.
 
Rather than using the Gamma prior on the scale and the LKJ prior on the correlation, it is more common in these sorts of hierarchical analyses to set the prior on $\Sigma$ to be the scaled inverse Wishart distribution \citep{Gelman2007}.  This choice is usually made for its conjugacy on Gaussian likelihoods and simplicity within Gibbs Sampling.  However, this distribution undesirably assumes a prior relationship between the variances and correlations \citep[see][for a review on priors for covariance matrices]{Alvarez2014}. In our sampling method, which we discuss in \autoref{sec:Model fitting}, conjugate priors are not necessary and, in fact, the combined scale-LKJ prior is more efficiently sampled and gives us control over the diagonal elements of $\Sigma$.

\subsection{Sample parameters}
The parameters that describe the properties of the ith cluster $\boldsymbol{x_i}$ are assumed to be drawn from the population distribution. We chose a centered parameterisation to draw cluster parameters from the population as the non-centered parameterisation \citep{Betancourt2013} suffered from biases and subpar performance as indicated by sampler diagnostics:
\begin{align}
	\boldx & \sim \mathcal{N}(\boldmu,\boldsymbol{L}\boldsymbol{L^\intercal})
\end{align}
where $\boldsymbol{L}$ is the Cholesky decomposition of $\boldSigma$.

This re-parameterisation is equivalent to drawing from a multivariate gaussian but is less computationally expensive since the covariance matrix is only decomposed once. It makes for more efficient sampling of the deformed regions of the parameter space commonly found in hierarchical inference problems. 
The probability of the parameters conditional on the global population takes the form of a multivariate gaussian distribution: 
\begin{equation}
	\mathrm{Pr}(\boldx|\boldmu, \boldSigma)=\prod_i \frac{1}{\sqrt{(2\pi)^{n}|\boldSigma|}} \exp\left[ -\frac{1}{2}	(\boldx_i-\boldmu)^{\intercal}\boldSigma^{-1}(\boldx_i-\boldmu)\right]
\end{equation}
where $n$=3 and
\begin{equation}
	\boldsymbol{x_i} = \left( \begin{array}{c}\rm\ln(M_{200}^{(i)}) \\ \ln(c^{(i)}_{200}) \\ \ln(1+z^{(i)}) \end{array} \right). \nonumber
\end{equation}

\subsection{Model fitting}\label{sec:Model fitting}
\begin{figure}
  \includegraphics{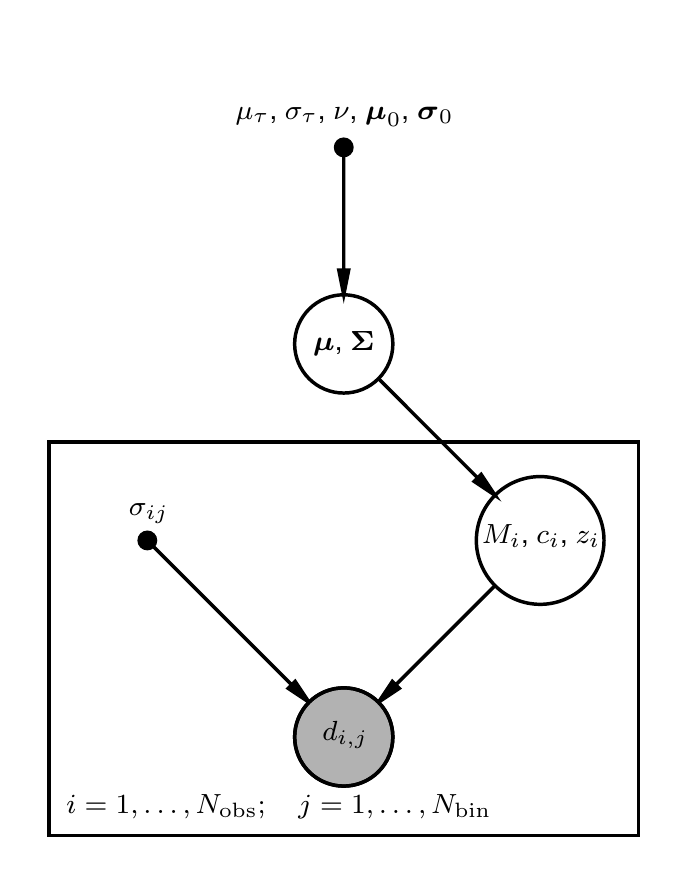}
  \caption{\label{fig:graphical-model} A graphical model of the
    relationships between terms in our posterior.  Filled ellipses
    indicate quantities that are observed and therefore conditioned-on in the
    analysis, while open ellipses contain parameters that are fit, and
    dots indicate fixed quantities that are not probabilistically
    modeled. At
    the top level, the (fixed) parameters controlling the hyper-priors
    influence the distribution of $\boldmu$ and
    $\boldSigma$. The parameters $\boldmu$ and
    $\boldSigma$ control the distribution of masses,
    concentrations, and redshifts of the individual clusters. The
    mass, concentration, and redshift of each cluster combine with the
    (fixed) observational uncertainties to control the distribution of
    the shear data.}
\end{figure}

The full posterior can be written as:
\begin{equation}
\mathrm{Pr}(\boldmu, \boldSigma, \boldx | \bold{d}) = \frac{  \mathrm{Pr}(\bold{d} | \boldx) \mathrm{Pr}(\boldx| \boldmu, \boldSigma ) \mathrm{Pr}(\boldmu)\mathrm{Pr}(\boldSigma)} { \mathrm{Pr}(\bold{d})}
\end{equation} 
where $\boldx$ are the individual cluster parameters and $\bold{d}$ are the data (shear profiles and spectroscopic redshifts). The likelihood is
\begin{multline}
  \label{eq:likelihood}
 \mathrm{Pr}(\bold{d} | \boldx) = \prod_i \frac{1}{\sqrt{(2\pi)}\sigma_{i,z}}  \exp\left[
   -\frac{\left(d_{i,z} - z\left( \boldx_i \right)\right)^2}{2
     \sigma_{i,z}^2} \right] \\ \times \prod_j \frac{1}{\sqrt{(2\pi)}\sigma_{i,j}} \exp \left[-\frac{(d_{i,j}-g(r_{i,j}, \boldsymbol{x_{i}}))^2}{2\sigma_{i,j}^2} \right],
\end{multline} 
where $d_{i,z}$ and $\sigma_{i,z}$ are the observed redshift and
associated uncertainty of the $i$th cluster, $d_{i,j}$ and
$\sigma_{i,j}$ are the observed shear and associated uncertainty of
the $i$th cluster in the $j$th radial bin, $z$ is the redshift
associated with parameters $\boldx$, and $g$ is the model
shear at the radius $r_{i,j}$ from the cluster centre. The model shear
is a function of the mass, concentration and redshift as computed
according to a NFW \citep{Navarro1997} density profile (see appendix \ref{a:NFWeqns}).
Regardless of the shear signal to noise ratio, we do not fix the
concentration to values from a mass-concentration relation; instead
information on the relationship between c-M flows through the
population distribution which is simultaneously fit to our data set.
We treat the quoted shear measurements as the fundamental data
product, and assume above in \autoref{eq:likelihood} that the sampling
distribution for the shear is Gaussian with width equal to the quoted
shear uncertainties. In reality, the fundamental data product of a
weak-lensing measurement is pixel-level images of background galaxies,
and the summary of this data by shear measurements induces a
distribution that is not Gaussian, but \autoref{eq:likelihood} is a
reasonable and computationally-efficient approximation. See
\citet{Schneider2015} for a discussion of hierarchically-modelled
pixel-level likelihood functions for shear maps; such models provide a
more accurate representation of the data but through a much more
complicated and expensive likelihood function. A graphical model of
our posterior appears in Figure \ref{fig:graphical-model}.

The Stan probabilistic coding language is used to implement inference on our problem with the R interface \cite{RStan-software2015}. Stan samples from posterior distributions using a Hamiltonian Monte Carlo (HMC) algorithm \citep{Neal2011, Betancourt2016}. HMC is a Markov Chain Monte Carlo (MCMC) sampling method where proposed states are determined by a Hamiltonian dynamics model.  This enables more efficient exploration of the parameter space and hence faster convergence in typical problems which is crucial for problems working in high dimensions.  

We run 4 chains with 1,000 warm-up samples followed by 1,000 monitored samples. Convergence is checked using trace plots, histograms of the tree depth and calculation of the Gelman-Rubin convergence criterion \citep[$\hat{R}<1.1$,][]{Gelman1992}. Sample bias is also checked by monitoring the number of divergences in a given sample. This diagnostic is specific to HMC, it indicates the number of numerical divergences occurred whilst sampling and is typical for regions of the parameter space that are hard to explore. Any number of divergences could suggest a bias in the posterior samples however it can be reduced by increasing the acceptance probability, or by re-parameterising the model.

\section{Predicting Future Data and Scaling Relations}\label{sec:predict-scaling} 

In order to use the results from the hierarchical model to predict parameters of future data, consider the following situation.  We observe or produce noisy estimates of (some of the) parameters of a previously unobserved system that we assume comes from the same population, and we now wish to use the population-level fitting to produce better estimates and/or predictions of parameters that we did not measure.  Let us assume that the observational uncertainties are Gaussian, described by a mean $\boldmu_0$ and a covariance matrix $\boldSigma_0$.  (If a parameter is un-observed, then we can set the corresponding diagonal element of the covariance matrix to $\infty$, indicating infinite uncertainty about its value). The true parameters of the system of interest are 
\begin{equation}
   \boldx_T =\{ M_T, c_T, z_T\},
\end{equation}
where we use $M_T, c_T, z_T$ as short-hand for the true underlying parameters $\ln M_{200}, \ln c_{200}, \ln(1+z)$.  Combining the results of the new observations with our population model results in a Gaussian distribution for the true parameters of the system with via, 
\begin{equation}\label{eq:ML_relation}
	\boldx_T \sim \mathcal{N}(\mu_T, \Sigma_T),
\end{equation}
where
\begin{eqnarray}
	\boldmu_T &=& \boldSigma_T (\boldSigma^{-1}\boldmu + \boldSigma_{o}^{-1}\boldmu_{o})\nonumber \\
	\boldSigma_T &=& \left(\boldSigma^{-1}+ \boldSigma_{o}^{-1} \right)^{-1}.
\end{eqnarray}
If we specialise to un-correlated measurements, then 
\begin{equation}
  \boldSigma_{o}^{-1} = \diag(\sigma_M^2, \sigma_c^2, \sigma_z^2)^{-1},
\end{equation}
a diagonal matrix with the uncertainties associated to each measurement.  

The posterior on the true parameters of an individual system is a normal distribution about the weighted mean of the population $\boldmu$ and the observable values $\boldmu_o$. The uncertainties are similarly dependent both on the population width $\boldSigma$ and the observational uncertainty $\boldSigma_{o}$. A small observable uncertainty will cause the parameter to be dominated by the observed value, whereas a large observable uncertainty will pull the parameter closer to the population estimate. This effect is particularly useful for measurements of low-signal to noise data. Where observables are missing, for example a measurement of a mass and redshift but no measurement of concentration, the hierarchical model can still be used as described above by setting $\sigma_c=\infty$.  In this particular case the estimate of $\boldmu_{T}^{c}$ would be weighted entirely by the population distribution at the appropriate values of $M$ and $z$.  We now proceed to derive Equation \ref{eq:ML_relation}.

Using Bayes theorem, the conditional distribution of the true parameters can be written:
\begin{align}
&\mathrm{Pr}(\boldx_T| \boldx_{o}, \boldSigma_{o}, \boldmu, \boldSigma) \nonumber \\
&\propto \mathrm{Pr}(\boldx_{o}|\boldx_{T}, \boldSigma_{o}, \boldmu, \boldSigma)\ \mathrm{Pr}(\boldx_{T}|\boldSigma_{o},\boldmu,\boldSigma)\nonumber \\
&\propto \exp\left[-\frac{1}{2}(\boldx_{o}-\boldx_{T})^\intercal \boldSigma_{o}^{-1}(\boldx_{o}-\boldx_{T})\right]\exp\left[-\frac{1}{2}(\boldx_{T}-\boldmu)^\intercal \boldSigma^{-1}(\boldx_{T}-\boldmu)\right] \nonumber \\
&\propto \exp\left[-\frac{1}{2}\left((\boldx_{o}-\boldx_{T})^\intercal \boldSigma_{o}^{-1}(\boldx_{o}-\boldx_{T})+(\boldx_{T}-\boldmu)^\intercal \boldSigma^{-1}(\boldx_{T}-\boldmu)\right)\right] \nonumber
\end{align}
The log posterior is thus proportional to a Gaussian distribution:
\begin{align}
&\mathcal{L}=-\frac{1}{2}\left((\boldx_{o}-\boldx_{T})^\intercal \boldSigma_{o}^{-1}(\boldx_{o}-\boldx_{T})+(\boldx_{T}-\boldmu)^\intercal \boldSigma^{-1}(\boldx_{T}-\boldmu)\right). \nonumber 
\end{align}
The posterior mean of $\boldx_t$ occurs at the maxima of the likelihood, where the derivative of $\mathcal{L}$ is 0. The posterior variance is the inverse of the negative second derivative of the $\mathcal{L}$. The first and second derivatives of the log-likelihood are
\begin{align}
&\frac{\partial \mathcal{L}}{\partial \boldx_{T}}=  \boldSigma_{o}^{-1}(\boldx_{o}-\boldx_{T}) + \boldSigma^{-1}(\boldmu-\boldx_{t}), \label{eq:L-first-deriv} \\
&\frac{\partial^2\mathcal{L}}{\partial \boldx_{T}^2}=  - \boldSigma_{o}^{-1}- \boldSigma^{-1}. \label{eq:L-second-deriv}
\end{align}
Setting $\partial \mathcal{L}/\partial \boldx_T = 0$ and solving for $\boldx_T \equiv \boldmu_{T}$ yields
\begin{align}
&\boldmu_T = \boldSigma_T (\boldSigma_o^{-1}\boldx_{o}+\boldSigma^{-1}\boldmu).
\end{align}
The variance $\boldSigma_T$ is
\begin{align}
&\boldSigma_T = -\left(\frac{\partial^2 \mathcal{L}}{\partial \boldx_{T}^2}\right)^{-1} = (\boldSigma_o^{-1}+\boldSigma^{-1})^{-1},
\end{align}
recovering the equations defined earlier.

\subsection{Scaling relations}
We can use the formalism above to derive scaling relations between the
parameters in our population model. A scaling relation is obtained
when two parameters are measured with zero uncertainty and a third is
unmeasured (i.e.\ with infinity uncertainty). For example, to compare
with existing c--M relations in the literature we can assume that we
measure mass and redshift perfectly and with no uncertainty i.e.
$\sigma_M = \sigma_z = 0$, $x_o^m=x_t^m$, $x_o^z=x_t^z$ and measure
concentration with infinite uncertainty i.e.
$\sigma_{o,c} \rightarrow \infty$, implying
$\Sigma_{o,22}^{-1} \rightarrow 0$ 
\begin{equation}
  \mu_T^c= \frac{\Sigma^{-1}_{12}}{\Sigma^{-1}_{22}}(\mu^m-x_T^m) +
          \frac{ \Sigma^{-1}_{23}}{\Sigma^{-1}_{22}}(\mu^z-x_T^z) +
          \mu^c. 
\end{equation}
If we replace $\mu_T^c$, $x_T^m$ and $x_T^c$ by $\ln(c_{200})$,
$\ln(M_{200})$ and $\ln(1+z)$ respectively
\begin{align}
\ln(c)=& \frac{\Sigma^{-1}_{12}}{\Sigma^{-1}_{22}}(\mu^m-\ln(M)) +
         \frac{ \Sigma^{-1}_{23}}{\Sigma^{-1}_{22}}(\mu^z-\ln(1+z)) +
         \mu^c. 
\end{align}
then we can rearrange into the familiar multiple regression form
\begin{align} \label{eq: cMrelation}
&\ln(c) = \alpha + \beta \ln(M)+ \gamma \ln(1+z) 
\end{align}
where
\begin{align}
&\alpha = \frac{\Sigma_{12}^{-1}}{\Sigma_{22}^{-1}}\mu^m +
  \frac{\Sigma_{23}^{-1}}{\Sigma_{22}^{-1}}\mu^z + \mu^c \nonumber \\ 
&\beta = -\frac{\Sigma_{12}^{-1}}{\Sigma_{22}^{-1}} \nonumber \\ 
&\gamma = -\frac{\Sigma_{23}^{-1}}{\Sigma_{22}^{-1}} \nonumber.
\end{align}

If we instead assume that mass and redshift are measured perfectly, we
derive a different scaling relation which is \emph{algebraically
  inequivalent} to the above relation because of the different
assumptions ($\sigma_M = 0$ versus $\sigma_c = 0$). In the event that
any of these quantities are actually measured, with associated
non-zero uncertainty, it is better to use the full formalism from
Section \ref{sec:predict-scaling} which takes into account measurement
uncertainty than to substitute into a scaling relation that ignores
it.

\begin{figure*}
\centerline{
\includegraphics[width=85mm]{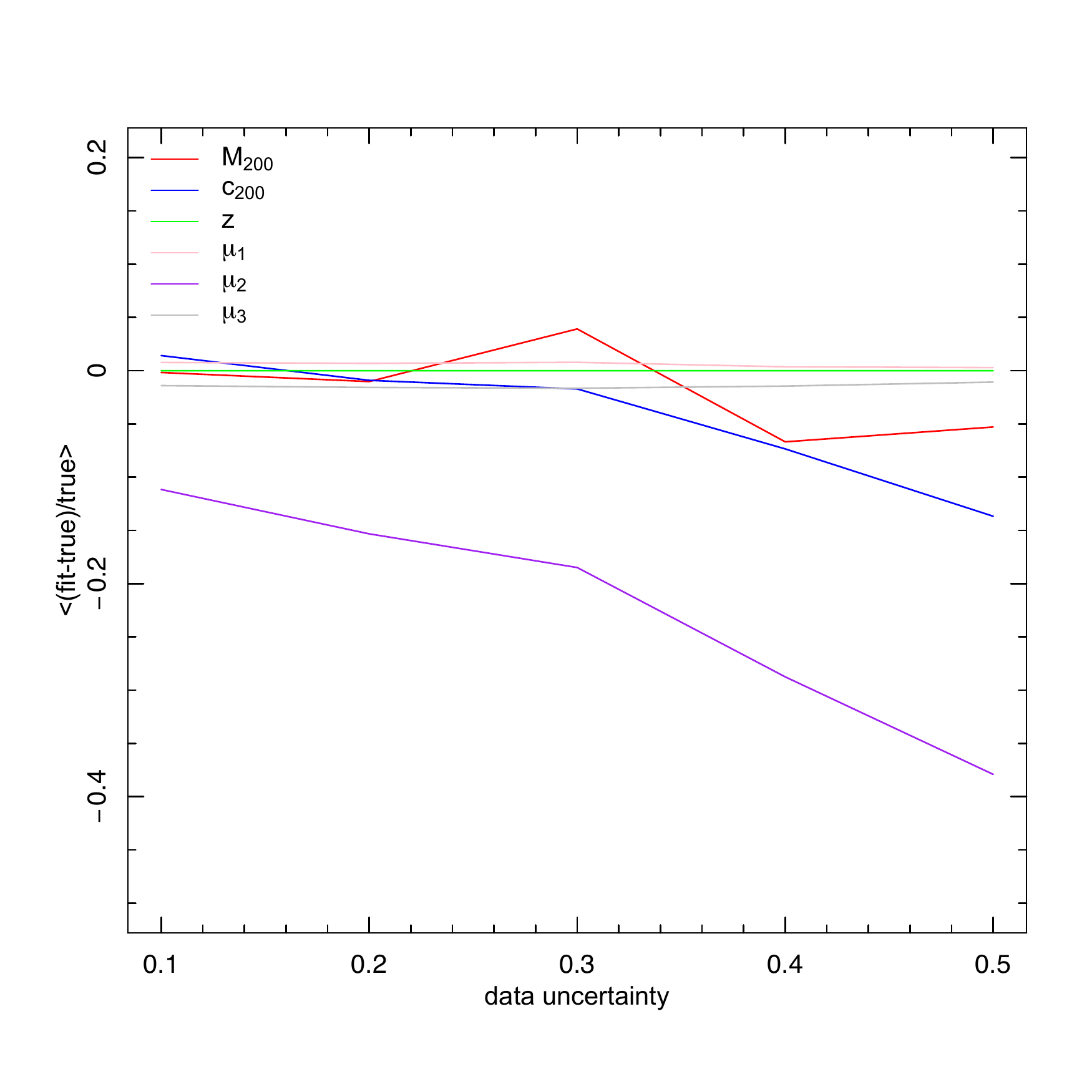}
\includegraphics[width=85mm]{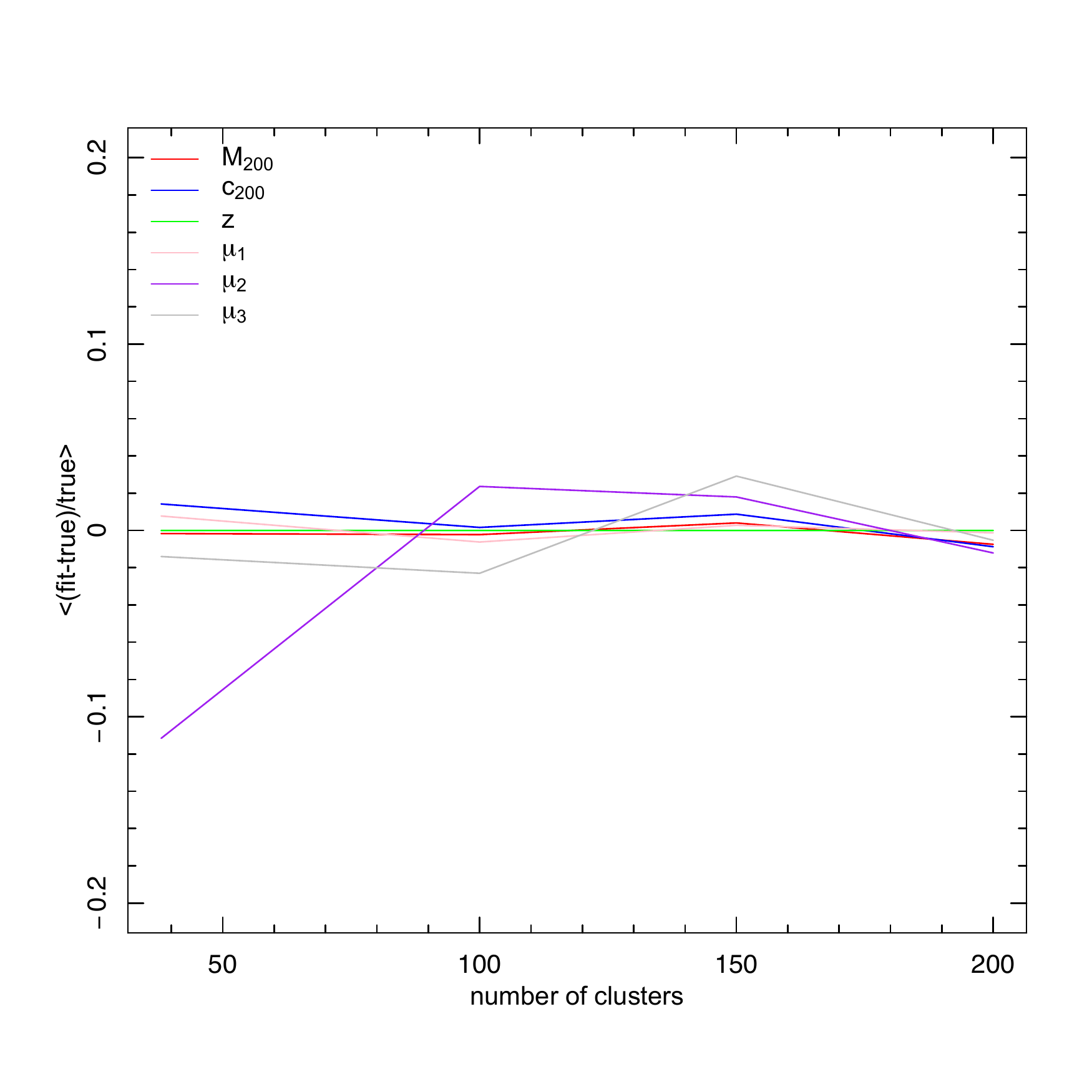}
}
\caption{The bias in measured parameters for toy simulations. The subscripts 1,2,3 on $\mu$ represent the $\ln(M_{200}[h_{70}^{-1}\ \rm M_\odot])$, $\ln(c_{200})$ and $\ln(1+z)$ population components respectively. Decreasing shear uncertainty (left) and increasing cluster sample (right) improves the ability to reproduce the truth. \label{fig: plot_compare_bias_TOY}
}
\end{figure*}
\begin{figure*}
\centerline{
\includegraphics[width=85mm]{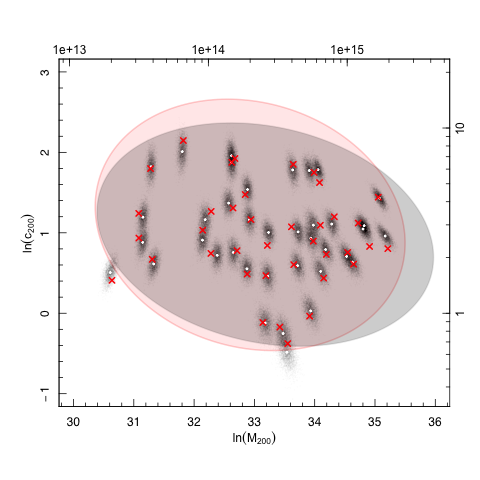}
\includegraphics[width=85mm]{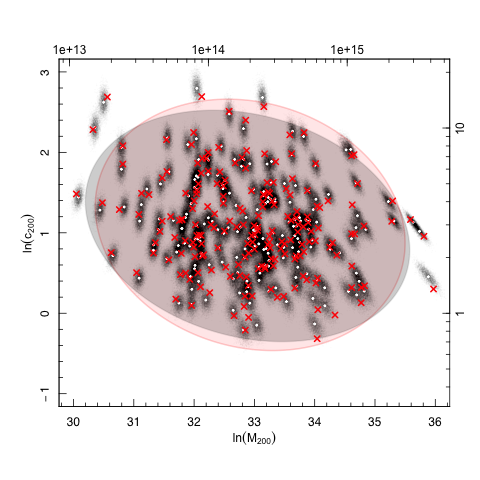}
}
\caption{The results of the toy simulations of 38 and 200 clusters with 10$\%$ error on shear. The true population is shown by the red ellipse (95$\%$ region) and the black ellipse is based upon the fitted values of mean population parameters (note that it does not take into account the posteriors on the population, only the point estimates $\overline{\mu}$ and $\overline{\Sigma}$). The red crosses indicate the input concentration and mass values and the white points show the mean of the fitted values.\label{fig: plot_compare_bias_ncl}
}
\end{figure*}

\begin{figure*}
\centerline{
\includegraphics[width=85mm]{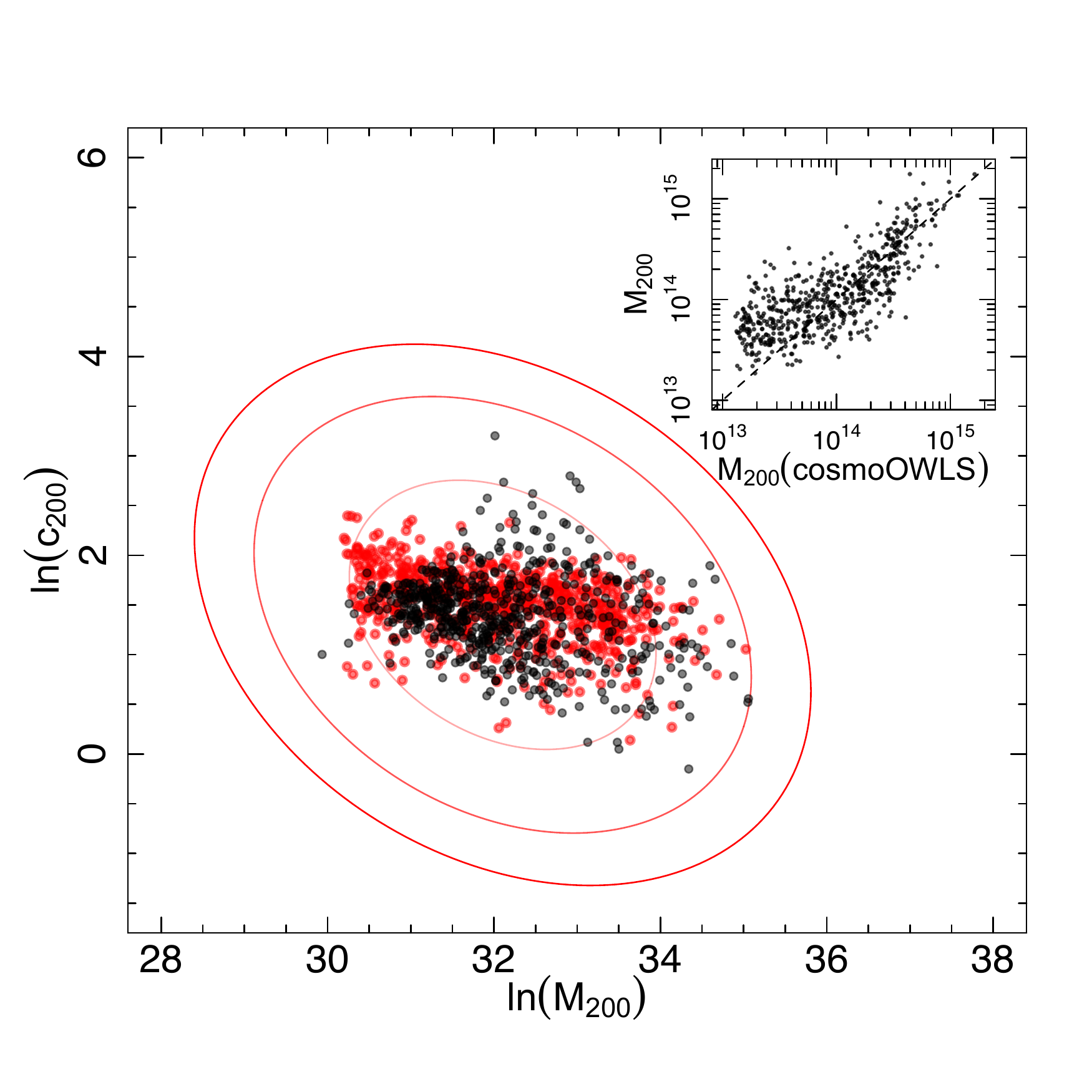}
\includegraphics[width=85mm]{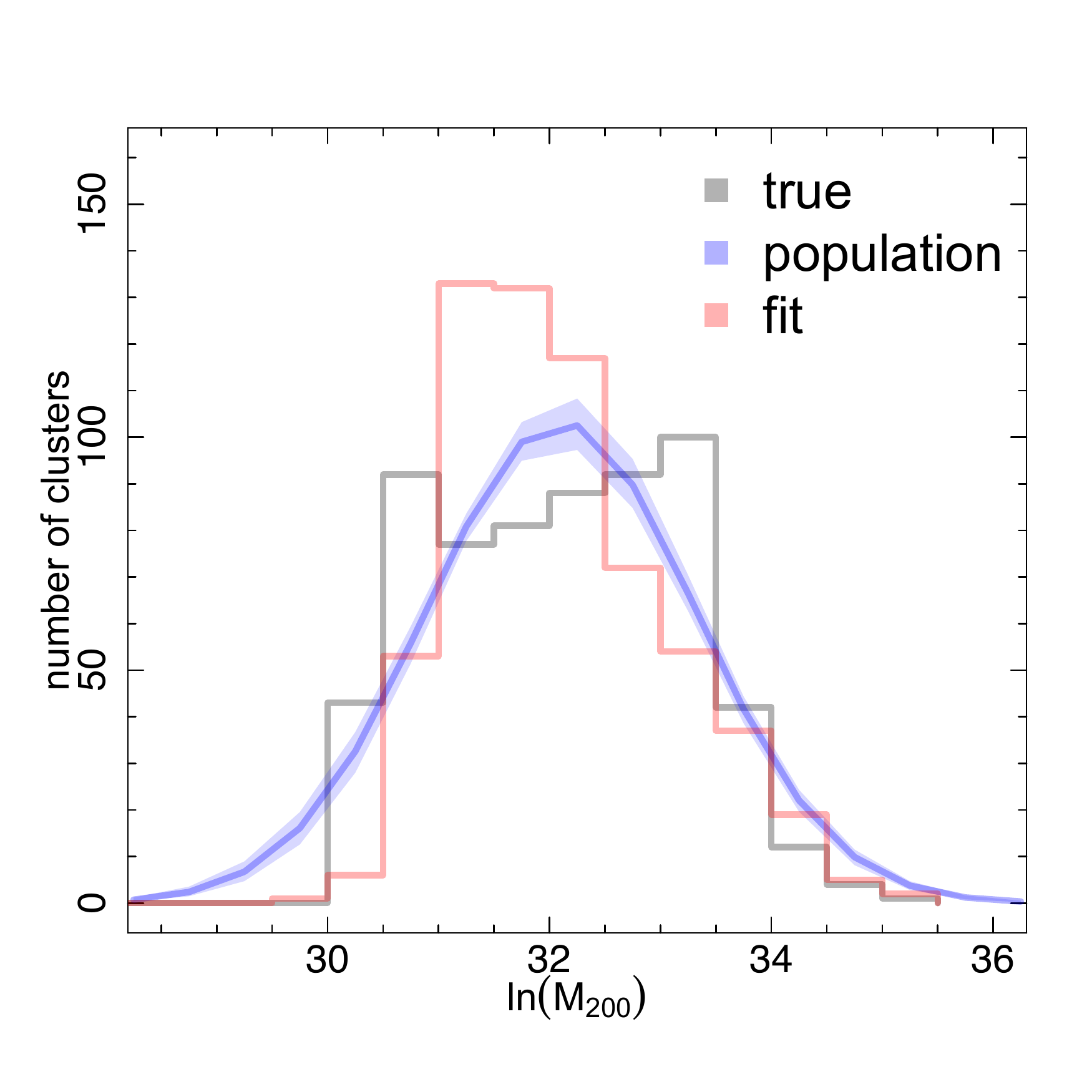}
}
\caption{{\sc Left:} True mass and concentration (red) of 632 cosmo-OWLS clusters at z=0.25 against fitted parameters in this work (black), the contour is derived from the point estimates of the population parameters $\boldmu$, $\boldSigma$ and inset is a comparison of masses. {\sc Right:} A histogram of the true cluster masses from simulations (black) and the comparison with the individual fitted masses in this work (red). The shrinkage can be seen clearly from the overestimation at $\ln(M_{200})$ = 32 in the fitted values and the underestimation at tails. From the point estimates of the fitted population parameters (blue) we recover a much better fit to the truth but this can be further improved if assume a population with the form of the cluster mass function rather than that of a gaussian. \label{fig: plot_cosmoowls}}
\end{figure*}

\section{Tests on simulated data}\label{sec: toydata}
We test our model on toy data by generating shear profiles for 38 clusters, each with 8 radial bins spaced equally in log. The masses, concentrations and redshifts are drawn from an arbitrary multivariate distribution of mean $\boldmu = \{\ln(2\times 10^{14}), \ln(3), \ln(1+0.3)\}$ and covariance $\boldSigma = \{(1.1,-0.1,0.05),(-0.1,0.4,0.05),(0.05, 0.05, 0.01)\}$. We note that the definition of the model in this form is not ideal since the $\ln(1+z)$ component in $\boldmu$ implies z can take negative values. It would be more natural to be expressed in the form $\ln(z)$, however this would then not allow the direct inference of the c-M relation in it's commonly expressed form. The (1+z) factor comes from the expansion factor of the Universe so is physically motivated. For this reason we make sure that the the cluster redshifts are positive. Trials of various levels of uncertainty on the shear measurements are made and we are able to recover parameters to within 2$\%$ uncertainty with the exception of the mean population concentration which is biased increasingly low as the uncertainty on the shear increases (\autoref{fig: plot_compare_bias_TOY}). None the less, the fitted values agree within the uncertainties and the bias improves significantly when increasing the sample size from 38 to 200 clusters (\autoref{fig: plot_compare_bias_ncl}) which is promising for the application of this work on upcoming large cluster surveys.

We also test our model on 632 clusters drawn from redshift slice z=0.25 of the cosmo-OWLS cosmological hydrodynamical simulations \citep{LeBrun2014}. We use the dark matter only run with WMAP7 cosmology, 5 source galaxies arcmin$^{-2}$ and 28$\%$ shape noise. Omitting the redshift component from our model we find, we can recover reasonably well the cluster parameters and population estimates (see \autoref{fig: plot_cosmoowls}a). In the high mass range, due to the high signal to noise of the data, the individually measured mass values give a good estimate of the true cluster number count. However we see that at lower mass this is not true and affects the predicted normalisation. On the other hand the estimates from the population shows good agreement at all mass scales (\autoref{fig: plot_cosmoowls}b). 

\begin{figure*}
\centerline{
\includegraphics[width=90mm]{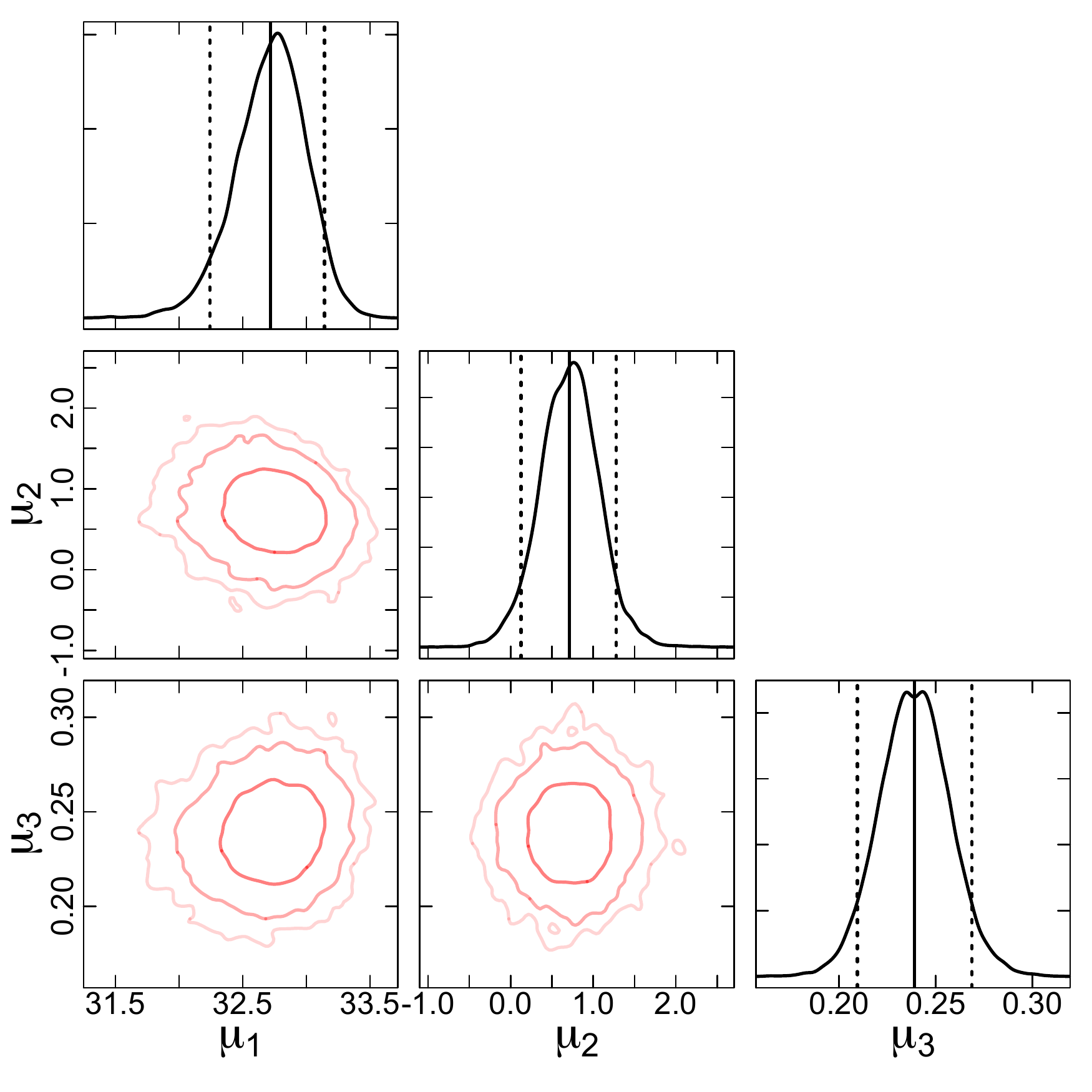}
\includegraphics[width=90mm]{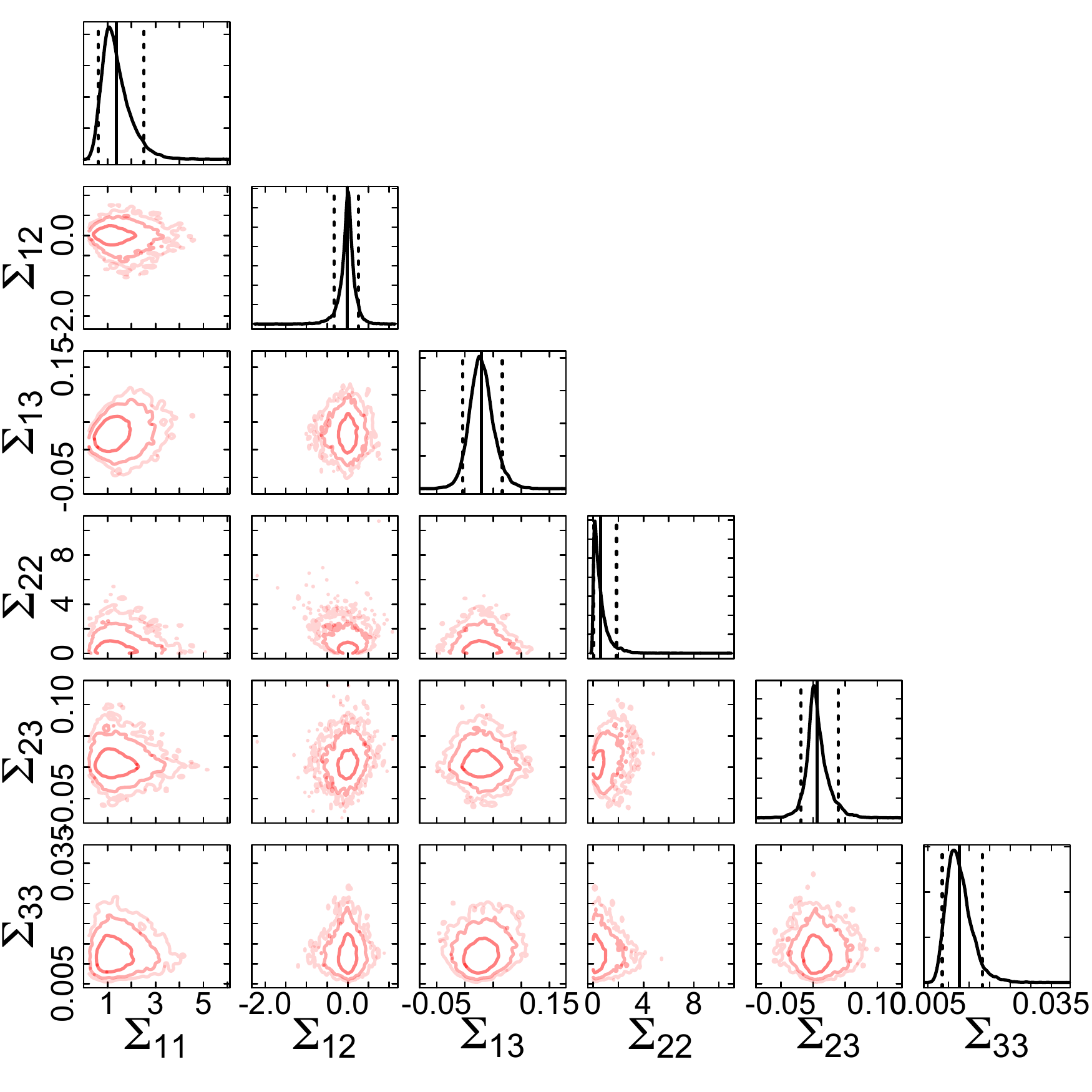}
}
 \caption{Posterior distributions of the 9 hyperparameters, where the subscripts 1,2,3 represent $\ln(M_{200}[h_{70}^{-1}\ \rm M_\odot])$, $\ln(c_{200})$ and $\ln(1+z)$ respectively. The red contours show 68, 95 and 99$\%$ confidence intervals, the histograms show the marginalised parameters with dashed vertical lines at 2$\sigma$. {\sc Left}: Global mean vector parameters {\sc Right}: Covariance matrix elements. \label{fig:plot_posterior_population}}
\end{figure*}

\begin{figure*}
\centering
\hspace{-0.8cm}
\includegraphics[width=65mm]{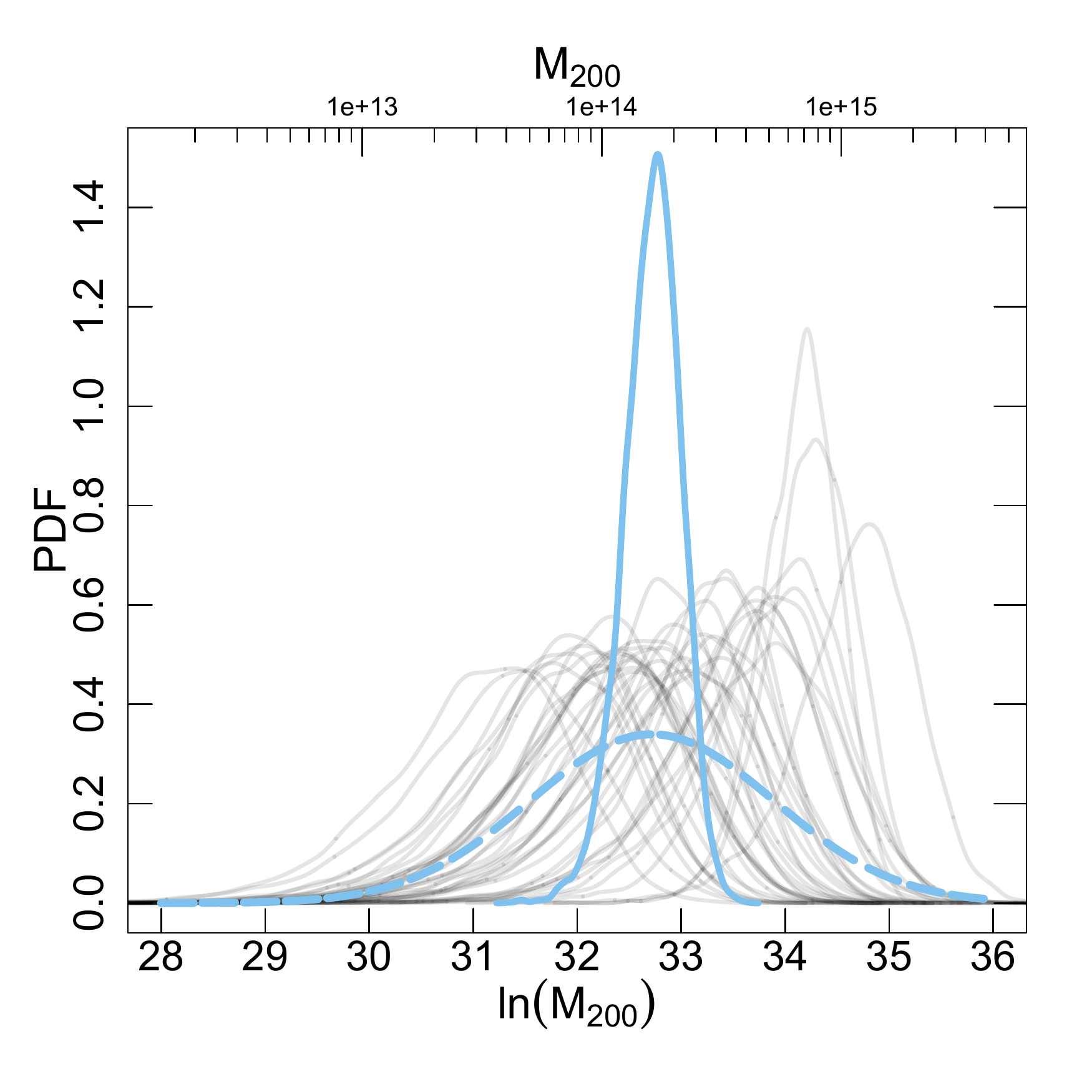}
\hspace{-0.8cm}
\includegraphics[width=65mm]{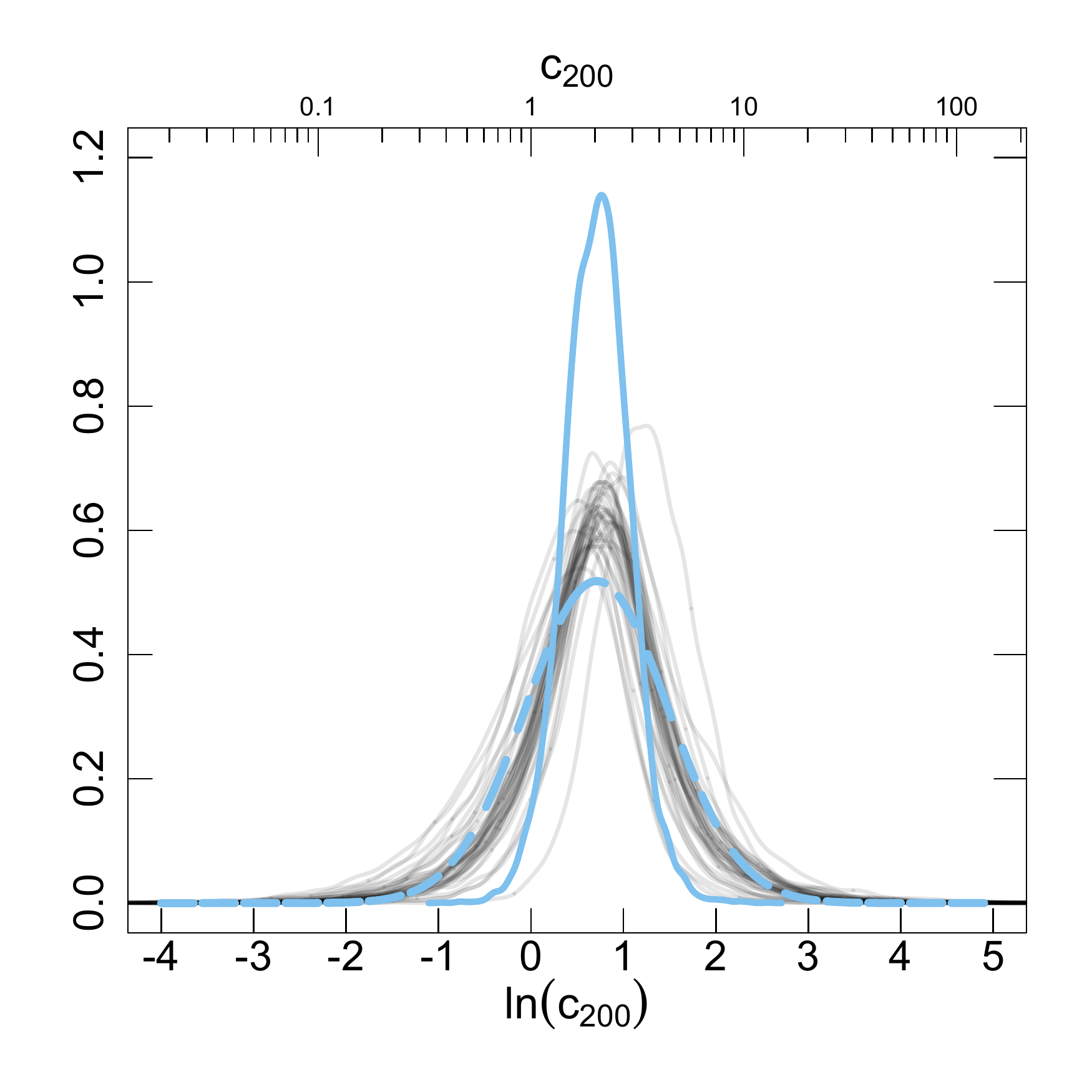}
\hspace{-0.8cm}
\includegraphics[width=65mm]{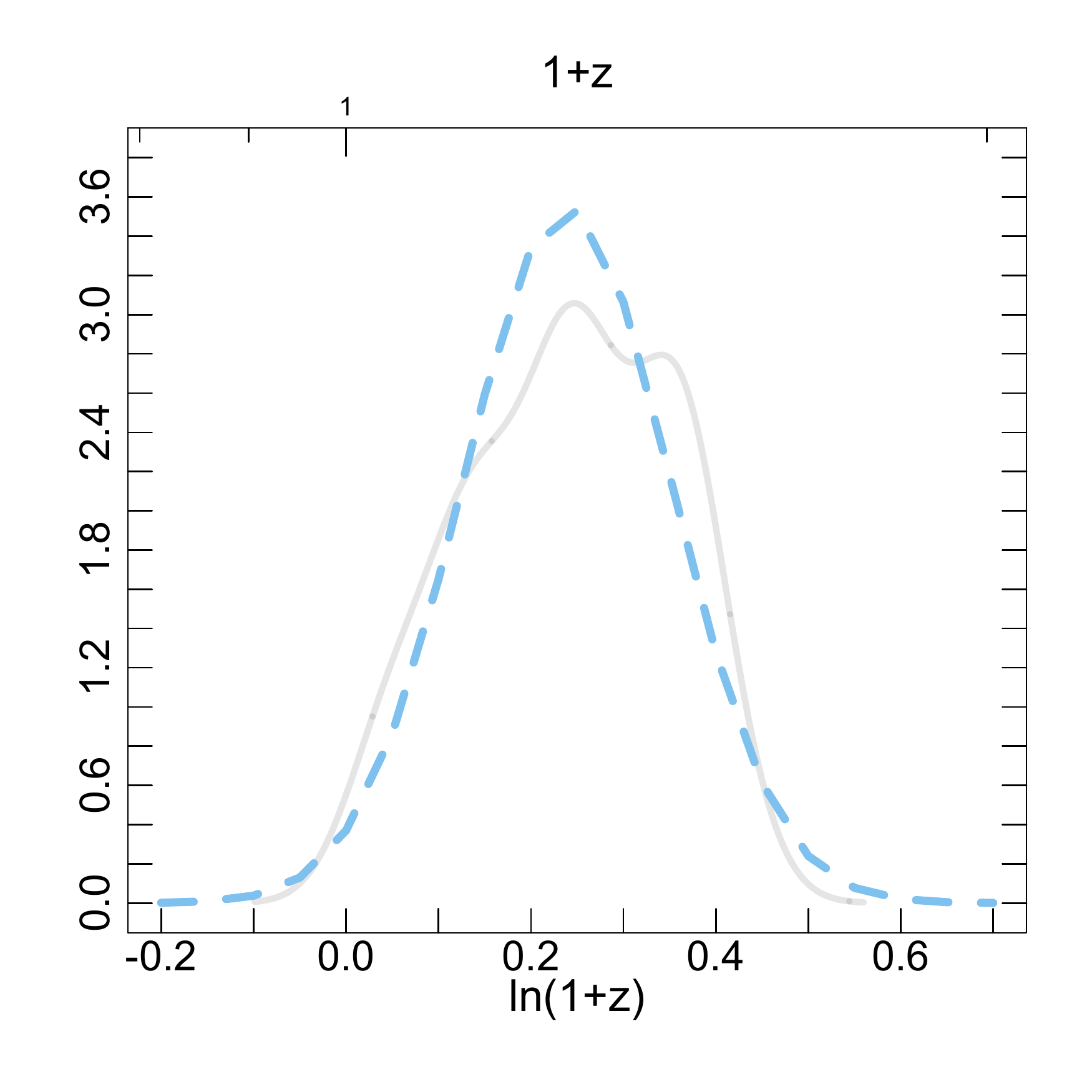}
 \caption{ {\sc Left $\&$ center:} Comparison of the posteriors for the population mean value (solid blue) and the posteriors for the individual clusters (solid grey) for mass and concentration respectively. A 1$\sigma$ deviation marginal centered on the mean of the population mean distribution (dashed blue). {\sc Right:} The z distribution of the population plotted as a gaussian centred on a mean and standard deviation obtained from the global mean vector and covariance matrix (dashed blue). It agrees well with the spectroscopic redshift distribution of the sample (shown as a gaussian kernel density estimate in solid grey). From this we can conclude that the data is able to constrain the individual cluster masses reasonably well, as the individual mass posteriors appear independent of the population mass posterior. On the contrary, the individual concentrations are completely dominated by the posterior of the population concentration, which implies that without the hierarchical model, individual cluster concentrations would not be possible. 
  \label{fig:plot_compare_mu_single}}
\end{figure*}

\section{Application to data}\label{sec:Results}
\subsection{Data}\label{sec:Data} 
We further apply our method to observational data from \cite{Lieu2016}. Here we provide a brief summary.

The sample consists 38 spectroscopically confirmed groups and poor clusters that lie at $0.05<z<0.6$ and span the low temperature range of $\rm T_{\rm 300kpc} \simeq 1- 5\,\rm  keV$\citep{Giles2016}. They are selected in X-ray to be the 100 brightest systems\footnote{XXL-100-GC data are available in computer readable form via the XXL Master Catalogue browser \url{http://cosmosdb.iasf-milano.inaf.it/XXL} and via the XMM XXL Database \url{http://xmm-lss.in2p3.fr}} and collectively lie within both the Northern field of the XXL survey \citep{Pierre2016} and the CFHTLenS survey\footnote{\url{www.cfhtlens.org}} \citep{Heymans2012,Erben2013}. The clusters are confined to z$<$0.6 due to limited depth of the CFHTLenS survey, this corresponds to a background galaxy cut of $\sim \rm  4\ arcmin^{-2}$. The sample is not simply flux-limited, the systems are selected based upon both count rate and extension \citep[see ][for details]{Pacaud2016}. 

We use shear profiles as computed in \cite{Lieu2016} that are distributed into 8 radial bins equally spaced on the log scale. They use a minimum threshold of 50 galaxies per radial bin which if not met, is combined with the subsequent radial bin. In the future we intend to extend the method to the full shear catalogue without binning. The errors on the shear are computed using bootstrap resampling with $10^3$ samples and incorporate large scale structure covariance.

All 38 clusters have spectroscopic redshifts, therefore we are able to use this information as data within the model. 

Our model applied to the XXL data set is 123-dimensional ($3 \times 38$ cluster parameters and $9$ population-level parameters). 

\subsection{Global estimates}
The posteriors of the hyperparameters approximately follow gaussian distributions (Figure \ref{fig:plot_posterior_population}). This justifies the use of the posterior mean and standard deviation as the estimator of the fits. For the global mean vector and covariance matrix these are
\begin{equation}
	\boldmu =  \rm \left( \begin{array}{c} 32.718 \pm 0.278 \\
					 0.711 \pm 0.357 \\ 
					 0.239 \pm 0.018 \end{array} \right), \nonumber
\end{equation}
\begin{equation}
	\boldSigma = \left( \begin{array}{ccc} 1.379 \pm 0.609 & -0.014 \pm 0.190 & 0.030 \pm 0.022 \\ 
					-0.014 \pm 0.190  & 0.593 \pm 0.644 & 0.007 \pm 0.018 \\ 
					0.030 \pm 0.022  	& 0.007 \pm 0.018  & 0.013 \pm 0.003 \end{array} \right). \nonumber
\end{equation}
A comparison between the population z distribution and the distribution of spectroscopic redshifts of the sample acts as a reassurance that the model is indeed working. We also compare the posteriors of $\mu_M$ and $\mu_c$ to the posteriors of M$_{200}$ and c$_{200}$ of the individual clusters (\autoref{fig:plot_compare_mu_single}). The individual concentration values are weakly constrained resulting in posteriors that are dominated by the population mean, whereas the individual masses are able to suppress the influence of the mean mass. This  demonstrates that independently, the individual clusters could not have constrained a concentration value.

 \begin{figure*}
\includegraphics[width=160mm]{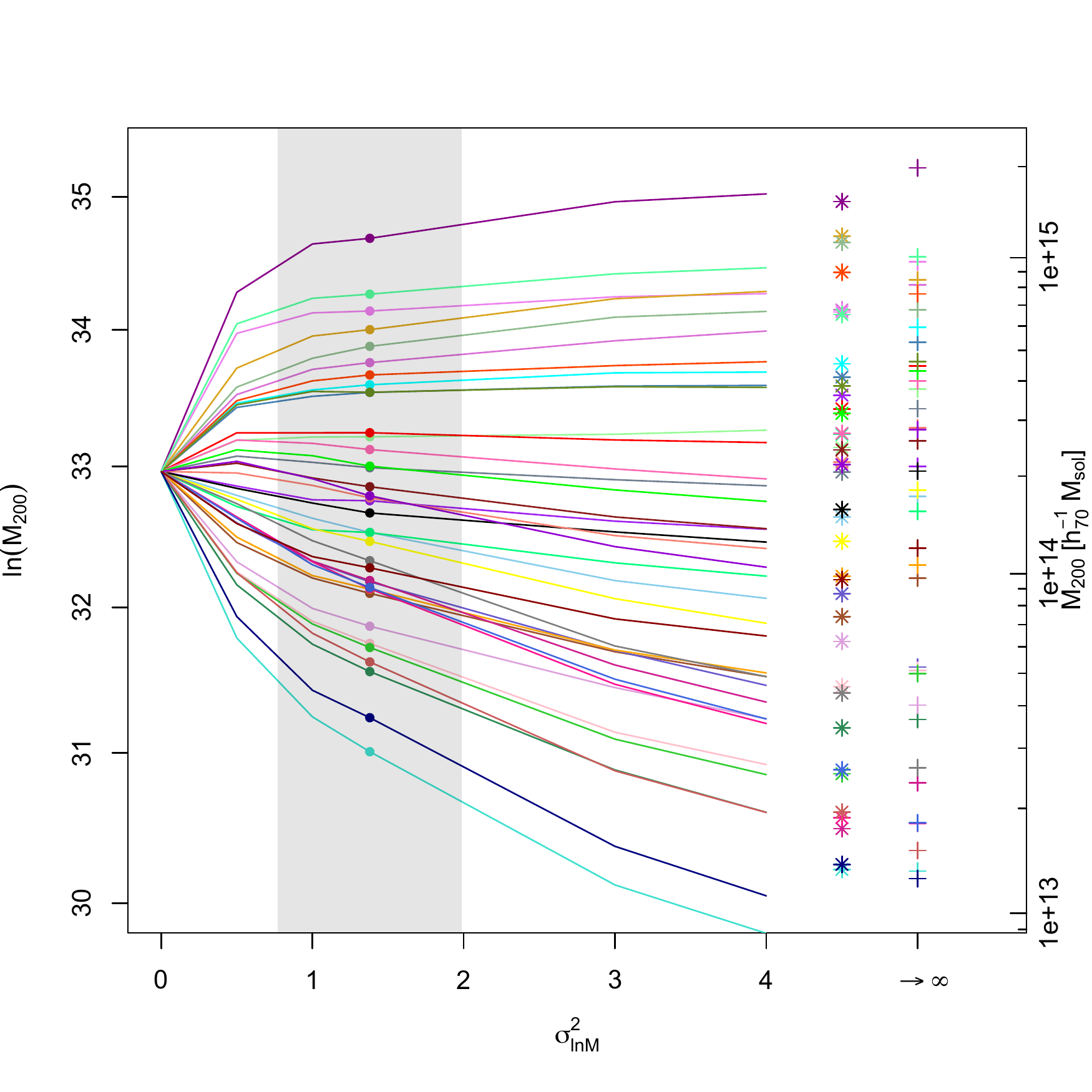}
 \caption{Individual galaxy cluster mass shrinkage estimates show the individual mass estimates shrink towards the population mean as $\sigma_{\ln M}^2$ decreases. Each cluster is represented as a different colour. The points show the fitted individual masses of clusters using the hierarchical method where $\sigma_{\ln M}^2$ is 1.38 and the shaded region is the 1$\sigma$ error. The stars and crosses are the individual masses following a non-hierarchical method  \protect\citep{Lieu2016}  where concentration is a free parameter and where concentration is fixed to the  \protect\cite{Duffy2008} c--M relation respectively.\label{fig:plot_shrinkage} }
\end{figure*}

\subsection{Mass estimates}
We find smaller masses to those computed independently from the individual shear profiles in \cite{Lieu2016} (\autoref{fig:plot_compare_old}). 

We calculate the weighted geometric mean between 2 mass estimates of n clusters as 
\begin{equation}
	\langle  M_1/M_2 \rangle = \exp\left( \frac{\sum_{i=1}^n w_i \ln\left(\frac{M_{1,i}}{M_{2,i}}\right)}{\sum_{i=1}^n w_i} \right).
\end{equation}
The weight is the expressed as a function of the error on the individual mass measurements ($\sigma_{M_{1}}, \sigma_{M_{2}}$). 
\begin{equation}
  \label{eq:weight-definition}
	 w_i = \frac{1}{\sigma_{\ln\left(\frac{M_{1,i}}{M_{2,i}}\right)}^2} = \left[\left(\frac{\sigma_{M_{1,i}}}{M_{1,i}}\right)^2+\left(\frac{\sigma_{M_{2,i}}}{M_{2,i}}\right)^2\right]^{-1},
\end{equation}
and the error we present on the mean is calculated from the standard deviation of 1000 bootstrap resamples. For an unbiased comparison we look at only non-upper limit measurements. With masses where the concentration is a fitted parameter we find that $\rm \langle M_{hierarchical}/M_{free} \rangle = 0.86 \pm 0.05$. In comparison to the masses assuming a fixed concentration following \cite{Duffy2008} c--M relation, the bias is $\rm \langle M_{hierarchical}/M_{Duffy} \rangle = 0.72 \pm 0.02$. However it is clear that it is not very informative to express the comparison in terms of a single number. The offset in mass is mass dependent, the hierarchical method measures significantly larger masses  for the upper limit/low mass systems as they are pulled towards the population mean. The comparison of the marginalised posterior distribution functions of the masses derived here and those derived independently with concentration as a free parameter, show reasonable agreement (\autoref{fig:plot_compare_old_mass_posteriors}). The obvious outliers are the low SNR objects which when treated individually show truncated posteriors at 1$\times 10^{13}\ \rm M_\odot$. This truncation arises from the implantation of a harsh prior boundary that is well motivated from the X-ray temperatures. For the same clusters, our masses all lie above $10^{13}\ \rm M_\odot$ but with very different values of mass, implying that even with a well motivated prior, the affect on mass can be significant. 

\citet{Smith2016} discussed alternative weight functions for
comparison of two sets of cluster mass measurements via a weighted
geometric mean calculation. They defined the weights for weak-lensing
masses in terms of $\sigma_M$, in contrast to our choice of
$\sigma_M/M$, arguing that for their sample the former definition was
more closely related to data quality than the latter, which tended to
up-weight more massive clusters. The \citet{Smith2016} sample spans a
smaller redshift and mass range than the sample that we consider here.
Therefore, issues relating to a mass-dependence of the weight function
are much less clear cut for our study than for \citet{Smith2016}. For
simplicity in this proof of concept study, we therefore adopt the more
conventional weight function given in Equation
\eqref{eq:weight-definition}.

\subsection{Shrinkage}\label{sec:shrinkage}
In the hierarchical model, mass is shrunk towards to the global population mean (Figure \ref{fig:plot_shrinkage}). In comparison to the individually fitted masses measured in \cite{Lieu2016}, equivalent to a population with mass variance $\sigma_{\rm \ln{M}}^2=\infty$, the hierarchical method is able to obtain better constraints on weakly constrained masses. Further, shrinkage estimates can be obtained by reducing the value of the relevant diagonal element of the global covariance matrix. As $\sigma_{\rm \ln{M}}^2 \rightarrow 0$, the mass estimates shrinks towards the global mean, which is equivalent to the mass obtained by stacking all clusters.
 
Assuming all clusters have a single mass value, whilst allowing concentration to be free we obtain a stacked mass estimate of $\exp(\ln M_{200}) =2.07 \pm \rm 0.79 \times 10^{14}\ \rm M_\odot$ with a global concentration value of $\exp(\mu_{\ln{c}})=1.78 \pm 1.72$

We can perform the same analysis for concentration, whilst allowing mass to be free we obtain a stacked concentration estimate of $\exp(\ln c_{200})=2.21\pm1.44$ with a global concentration value of $\exp(\mu_{\ln{M}})=1.62\pm 1.01\times 10^{14}\ \rm M_\odot$

A simultaneous fit for a single stacked mass and concentration, results in 1.91 $\pm 0.70\rm \times 10^{14}\ \rm M_\odot$ and 1.60$\pm 1.16$ respectively. Hence both parameters are in agreement within the errors either based on stacking only on either one of those parameters or both. The constraints on mass are stronger than concentration as expected due to the difficulty in measuring the latter.

The global means for the hierarchical fit were $\exp(\mu_{\ln{M}}) = 1.62 \pm 1.04 \times 10^{14} \rm M_\odot$ and $\exp(\mu_{\ln{c}}) = 2.04 \pm 1.68$  Although within the errors these results are consistent with the shrinkage estimates, the mean mass is slightly smaller and the mean concentration slightly larger. Simple stacking is a more severe constraint on M--c; blindly stacking clusters together can lead to incorrect mass estimates. In particular, our constraint on concentration is poor and therefore the mass estimates are not too sensitive to the concentration. More data is required to achieve a reliable estimate of the mean concentration of the population.

 \begin{figure}
\includegraphics[width=85mm]{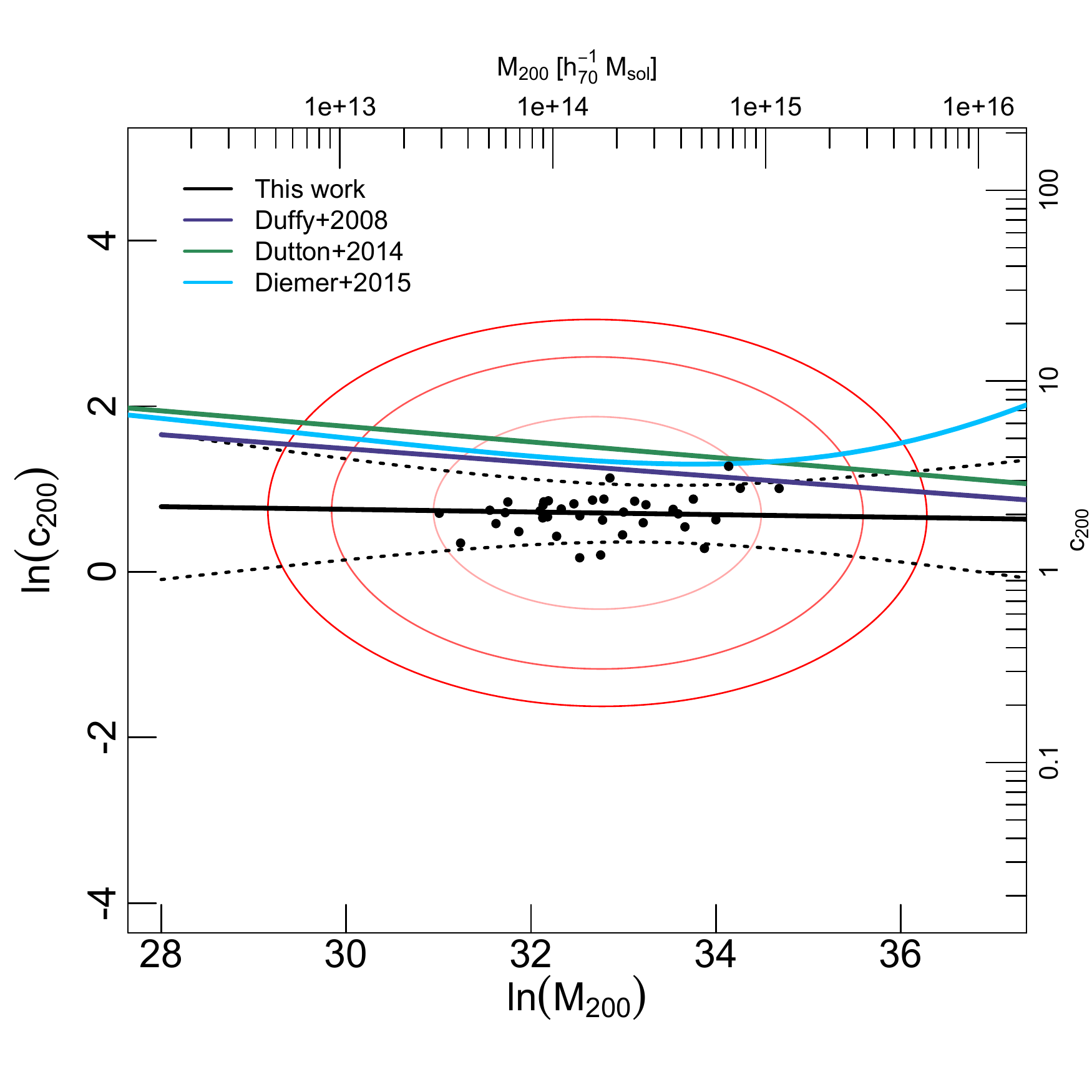}
 \caption{Concentration -- mass relation. $c$ values are computed for all $M$ values in the range using the equation \ref{eq: cMrelation} for each pair of $\boldmu$ and $\boldSigma$ sampled. The mean and 1$\sigma$ uncertainty is shown as the solid black and dotted lines respectively. The fitted covariance and mean of population concentration and mass shown by red contours of 1, 2 and 3$\sigma$ confidence and therefore appear mis-aligned from the fit. For comparison we plot the c--M relations from \protect\cite{Duffy2008} (solid purple line), \protect\cite{Dutton2014} (solid green line) and \protect\cite{Diemer2015} (solid blue line) are shown at our population mean redshift z=0.27. The black points are the mean of the individual log parameters.\label{fig:plot_c_M_relation}}
\end{figure}

\subsection{Mass -- concentration relation}\label{sec:Mcrelation}
Using \autoref{eq: cMrelation} we obtain mean values of $\alpha$ = $1.09^{+5.11}_{-2.85}$, $\beta$ = $-0.02^{+0.11}_{-0.36} $, $\gamma$ = 0.59$^{+1.54}_{-0.90}$ (Figure \ref{fig:plot_c_M_relation}). Note that the majority of the individual masses lie within 1$\sigma$ since it is based not on the means of the masses but the posteriors. Here the 1 $\sigma$ ellipse encompasses a third of the combined individual posteriors. 
 
We find concentrations that are typically smaller than \cite{Duffy2008} and \cite{Dutton2014} though the slope of our relation is compatible. We note that with the quality of the data, we are unable to constrain concentration leading to large uncertainties on our regression parameters. Our data marginally agrees with a weak anti- correlation between concentration and mass that is expected from mass accretion history theory \citep{Bullock2001}. Low mass groups formed in early times when the mean density of the Universe was larger, allowing concentrated cores to form. Massive clusters formed later on through the accretion of groups. In the literature, the concentration--mass relation is primarily estimated using numerical simulations where the concentration parameters are known exactly. Where a c--M relation has been measured from observations, studies have relied on high signal-to-noise clusters or stacking multiple clusters together to obtain a concentration estimate. We have already seen from the shrinkage estimates that stacking can cause overestimation of concentration. We note that the individual measurements of the $\ln\langle c_{200}\rangle$, $\ln\langle M_{200}\rangle$  are consistent with the higher values of concentration seen in the literature, however our assumption that these parameters are log normally distributed means that the correct values should be taken as $\langle\ln c_{200} \rangle$ and $\langle \ln M_{200}\rangle$ where the latter gives a result that is closer to the posterior peak of both Pr(x) and Pr($\ln(x)$) where x is c$_{200}$ and M$_{200}$. Due to the large uncertainties, our results are not able to rule out higher concentrations (\autoref{fig:plot_compare_mu_single}b). The c--M relations taken from the literature lie comfortably within the contours of the population mean and covariance.

\section{Discussion}

\subsection{Tests on priors}\label{sec:testpriors}
We test the influence of the priors on the toy data set. Recall that $\mu\sim\mathcal{N}(\mu_0, \sigma^2=1)$ where $\mu_0=(32,1,0.3)$. For the 38 clusters and 0.1 shear data uncertainties, the estimated population mean was $\bar{\mu}=(33.18 \pm 0.19, 0.98 \pm 0.09, 0.26 \pm 0.01)$. We vary the values of $\mu_0$ and find that the weakly informative prior does not affect the estimated population mean $\mu$ (see \autoref{tab:testpriors}). For $\mu_0[1]$ = (30, 31, 32, 33, 34) the mean of the posterior samples is $\langle \mu_1 \rangle = 33.18 \pm 0.19$ and for $\mu_0[2]$ = (-1, 0, 1, 2) we obtain $\langle \mu_2 \rangle = 0.97 \pm 0.09$.

Testing the prior influence on the observational data we again find that it does not affect the mean population mass and only weakly influences the estimation of the mean population concentration. Recall the measured population mean was $\bar{\mu}=(32.72 \pm 0.28, 0.71 \pm 0.36, 0.24 \pm 0.01)$. For the same varying values of $\mu_0[1]$ as above we obtain $\langle \mu_1\rangle = 32.71 \pm 0.13$ and for $\mu_0[2]$, we obtain $\langle \mu_2 \rangle = 0.63 \pm 0.16$. For the observational data, at the extreme hyper prior variants we get handful of divergences ($<$10/4000) even with a very high acceptance (0.999), as the prior extends further into the unstable regions of the parameter space. Given that the divergences remain sparse, we expect that the results to not be strongly affected.  

\subsection{Comparison to literature}

The concentration--mass relation is still a topic of interest since the regression parameters throughout literature vary significantly and observationally, the uncertainties are large \citep{Sereno2015}. Observation based c--M relations tend to rely on stacking analyses or samples of high signal-to-noise systems. We neither stack our lensing signals nor limit our sample to a signal-to-noise threshold since this may lead to a bias. Here we discuss and compare our results to the literature. 

Our data on average show lower concentration values compared to the \cite{Duffy2008} c--M relation which is known to be lower than many other simulation based c--M relations  \citep[e.g.][]{Okabe2013, Dutton2014}. However their relation assumes $WMAP5$ cosmology (we use $WMAP9$), and the inferred cosmology is known have a non-negligible effect on concentration \citep{Maccio2008}. Further, c--M relations based on numerical simulations tend to lower normalisations in comparison to observational samples. This could be due to selection effects or the physics included in the simulations. 

Using cold dark matter simulations based on $Planck$ cosmology \cite{Dutton2014} find a c-M relation whose evolution is not described by others in the literature (see their figure 11). Our data suggest a slight positive redshift evolution however with large uncertainties that are fully consistent with little or no evolution. Like many simulation based studies \citep{Klypin2014}, they find the Einasto density profile to be a better model for dark matter haloes in comparison to the NFW profile, however the significance is more pronounced for massive systems. \cite{Gao2008} find that the Einasto profile improves the sensitivity of concentration estimates to the radial fitting range in particular for stacked clusters. To implement this model however would require the introduction of 5 extra hyper parameters, and 38 parameters.  More importantly baryon physics is expected to play a more significant role in low mass systems which are not included in these simulations. Feedback affects both the normalisation and slope of the c--M relation by simultaneously decreasing the mass and increasing the scale radius, and massive neutrino free streaming can further lower the amplitude by reducing the mass (Mummery et al. in prep).

\cite{Okabe2013} have shown that the NFW profile fits well to the observations of stacked weak lensing data. Our method imposes a quasi-stacking so NFW may be appropriate. Compared to \cite{Duffy2008} and \cite{Dutton2014} our relation are $41\%$ and $49\%$ systematically lower respectively although only with a significance of 1.04 and 1.40$ \sigma$ lower. Our low concentration is consistent within the uncertainties with the literature \citep{Sereno2013}, however they consider higher redshift clusters (0.8$<$z$<$1.5). Such low concentrations are typical for halos undergoing rapid mass accretion and tend to be less well fit by the NFW profile.

Concentration is also correlated to the halo mass accretion history which in turn depends on the amplitude and shape of the initial density peak. In search for a universal halo concentration, \cite{Diemer2015} instead fit concentration to peak height $\nu$, their relation is also higher than ours however they find an upturn at high mass ($\nu$) scales. This flattening and upturn of the c--M relation is also found in other studies \citep{Bullock2001, Eke2001, Prada2012} and is attributed to there being more unrelaxed halos at higher mass. In our data, we too observe an upturn at higher mass scales however the low SNR of low mass clusters will have larger concentration errors so it is difficult to confirm this with the current small cluster sample. 

\cite{Bahe2012} use mock weak lensing observations based on numerical simulations to study the bias and scatter in M and c. They find that substructure and triaxiality can bias the concentration low ($\sim 12\%$) with respect to the true halo concentration, with the effect of substructure being the dominant effect. It can also lead to large scatter whilst having a much smaller effect on M$_{200}$. We expect this effect to be small on our sample because substructure and triaxial halos are more characteristic of massive clusters. 

Recently, \cite{Du2015} use 220 redMaPPer \citep{Rykoff2014} clusters with overlap with CFHTLenS to calibrate an observations c--M relation without stacking. They find a relation consistent with simulations but with large statistical uncertainties. Their clusters are slightly more massive then ours (M$_{200} \sim 10^{14} -10^{15}\ \rm M_\odot$) none the less their results suggest that the c--M relation is highly sensitive to the assumed prior (their Figure 6.). They find that dilution by contaminating galaxies and mis-centering can negatively bias the concentration values, we expect the latter to be more important in this work since we use spectroscopic redshifts and a conservative background selection, but our shear data is centred on the X-ray centroid.  By including priors based on richness and centring offset in their model, their results change significantly. Consequently we expect our c-M relation to change in the future with the inclusion of other cluster properties. 

It is important to note that, like mass measurements, concentration values observed using different methods and definitions may vary. Concentrations derived from weak gravitational lensing, strong lensing and X-ray are yet to reach agreement \citep{Comerford2007}.  

Possible reasons that the low normalisation of our c-M relation include the assumed cosmology, internal substructure, halo triaxiality or galaxy formation related processes that expel baryons into the outer regions of the halo resulting in a shallower density profile \citep{Sales2010, vanDaalen2011}. Also, as noted above, neutrinos can also lower the amplitude. Centre offset is degenerate with the normalisation of the c-M relation and neglecting any mis-centering could bias concentrations low \citep{Viola2015}. In our work we centre shear profiles on the X-ray centroids which may not trace the centre of the dark matter halo as well as the BCG but this should be accounted for since the inner radius of 0.15 Mpc is excluded when fitting the NFW model.

Another important point regards the imposed multivariate gaussian model and how well it fits the data. \autoref{fig:plot_compare_mu_single} shows that the posteriors of the individual cluster concentrations agree well with the gaussian prior however the masses appear more constrained by the single gaussian fit. A mixture model of 3 or more gaussians may be a better prior for the mass, however the additional flexibility introduced will also affect our ability to constrain concentration. For this work there is no reason to believe that the clusters do not originate from the same underlying population since they are selected in the same way, however in the future if external samples are to be included then the addition of further gaussians will be more important.

\cite{Sereno2015b} and subsequent papers in the $\textit{CoMaLit}$ (COmparing MAsses in LITerature) series, compare and apply methods to analyse cluster masses and scaling relations in a homogenous way whilst attempting to taking into account sample calibration issues that may lead to discrepancies in mass and scaling relations. Since in this paper we only explore mass, concentration and redshift whilst the clusters are selected on X-ray count rate, we leave observation biases to a future paper that will include observables such as X-ray luminosity.

\section{Conclusion}\label{sec:Conclusion}
We have developed a hierarchical model to infer the population properties of galaxy groups and clusters, and present a method for its correct usage to estimate of unknown parameters of additional clusters that is superior to the ad hoc scaling relation. Nevertheless familiar scaling relations can also be extracted. We apply the method on toy data, hydrodynamical simulations and observational data. Using this model we are able to obtain weak lensing mass estimates of individual clusters down to 1$\times10^{13}\ \rm M_\odot$ without the need for harsh prior boundaries and assumptions about the concentration, even when the signal-to-noise is low. Below is a summary:
\begin{enumerate}
\item We test the model on simulated toy data and find that the agreement with the true cluster mass and concentration is good and can further improve with increasing sample size and/or data uncertainties. Our tests on realistic weak lensing measurements from hydrodynamical simulations similarly show promising agreement.
\item We then apply the method on a small sample of 38 low mass groups and clusters from the \cite{Lieu2016}. Using this hierarchical method we are able to achieve better constraints on both mass and concentration without the compromise of upper limit measurements or the use of an external concentration--mass relation. This eliminates the bias introduced from calibrating with information derived from a sample that may not be representative of our systems. What's more the concentrations used in \cite{Lieu2016} are derived from dark matter only simulations, the missing physics could invoke differences from observations. The tests on the simulations is promising for the extension of this work to the full XXL sample ( $\gtrsim$ 600 galaxy groups and clusters).
\item \cite{Eckert2016}'s study on the XXL-100-GC galaxy clusters in the XXL survey find a very low gas fraction that requires a hydrodynamical mass bias of $M_X/M_{wl}=0.72^{0.08}_{0.07}$ to reconcile the difference. We measure masses on average 28$\%$ smaller compared to the mass estimates from \cite{Lieu2016} which may be able to resolve the issue. We note however that at the low mass end we measure higher masses compared to \cite{Lieu2016}, which would drive those gas fractions even lower. 
\item The mean population cluster mass and concentration are measured to be $\mu_M=1.62\pm1.04\times10^{14}M_\odot$ and $\mu_c=2.04\pm1.68$. The shrinkage of individual masses towards the population mean suggest that hierarchical modelling has a larger effect on the low mass systems where the signal-to-noise ratio is low. Tests with shrinkage of parameters suggest that blindly stacking clusters for mass and concentration can bias the estimated value of the population mean. Parameterising a single concentration whilst allowing mass to be free results in a concentration that is biased high compared to the population mean by 8$\%$. Stacking both concentration and mass to a single value on the contrary results in a positive mass bias of 18$\%$ and negative concentration bias of 22$\%$. This is worrisome for studies that rely on single concentrations for mass estimation those that blindly stack large samples of clusters. 
\item We estimate the concentration--mass relation from the underlying population obtaining a result which within the uncertainties is consistent with the literature. We are able to recover the weak anti-correlation between concentration and mass, however we find the data suggests much lower concentrations than those previously measured in observations and simulations. 
We attribute this to the fact that observation based c--M relations rely on stacking analyses which we do not use and as stated previously stacked concentration estimates tend to be biased high. Our c--M relation suggests an evolutionary dependance, however within the errors is not able to rule out no evolution. 
\end{enumerate}
Our method can be easily modified to incorporate more population parameters such as X-ray temperature, luminosity, gas mass etc. The additional cluster information will help to improve the constraints on mass predictions. In the future we hope to extend to cosmological inference by implementing a more accurate function to describe the population of clusters, namely convolving the true selection function with the cluster mass function. When the weak lensing data for XXL-south clusters becomes available we will be able to incorporate the additional systems to improve constraints on our model as well as other cluster samples in the literature \citep[e.g. see][]{Sereno2015}. This work will be important for current wide field surveys (such as DES, KiDS etc) where the data may be limited by the shallow survey depth, and for future big data surveys (e.g. Euclid, LSST, eRosita) who will need more efficient ways to deal with processing the predicted quantities of data whilst extracting the maximum amount of information from them.

\section*{Acknowledgements}
 
We thank Catherine Heymans, Gus Evrard and Jim Barrett for helpful discussions. We thank David van Dyk for a colloquium on shrinkage that inspired this work.  We are grateful to the CFHTLenS team for making their shear catalogue publicly available. ML acknowledges a Postgraduate Studentship from the Science and Technology Facilities Council and a ESA Research Fellowship at the European Space Astronomy Centre (ESAC) in Madrid, Spain. MS acknowledges the financial contribution from contracts ASI-INAF I/009/10/0, PRIN-INAF 2012 `A unique dataset to address the most compelling open questions about X-Ray Galaxy Clusters', and PRIN-INAF 2014 1.05.01.94.02 `Glittering Kaleidoscopes in the sky: the multifaceted nature and role of galaxy clusters'. 

\bibliographystyle{mnras}
\bibliography{HierarchicalMassModel}

\onecolumn
\appendix
\setcounter{figure}{0}  
\setcounter{table}{0}

\section{NFW density profile model}\label{a:NFWeqns}
The 3D NFW density profile is defined as,
\begin{equation}
\rho_{NFW}(r) = \frac{\rho_s}{(r/r_s)(1+r/r_s)^2},
\end{equation}
where the central density is,
\begin{equation}
\rho_s= \frac{200}{3} \frac{\rho_{cr}c^3}{\ln(1+c)-c/(1+c)}.
\end{equation}
Here $\rho_{cr}=(3H^2(z))/(8\pi G)$ is the critical density of the Universe at redshift $z$, where $H(z)$ is the Hubble parameter and $G$ is Newton's gravitational constant.  
We fit our data to the reduced gravitational shear 
\begin{equation}
g_{NFW} = \frac{\gamma_{NFW}}{1-\kappa},
\end{equation}
where the convergence can be expressed as the ratio of the surface mass density and the critical surface mass density $\kappa=\Sigma/\Sigma_{cr}$ and 
\begin{equation}
\Sigma_{cr}=\frac{c^2}{4\pi G}\frac{D_S}{D_L D_{LS}}
\end{equation}
where $c$ is the speed of light and $D_S, D_L$ and $D_{LS}$ are the angular diameter distances between the observer-source, observer-lens and lens-source respectively.
The shear is the difference between the mean surface mass density and the surface mass density 
\begin{equation}
\gamma_{NFW}=\frac{\overline{\Sigma}-\Sigma}{\Sigma_{cr}} 
\end{equation}
To obtain $\Sigma$ and $\overline{\Sigma}$ we integrate the 3D density profile along the line of sight $l$,
\begin{eqnarray}
\Sigma(x) &=& 2 \int^\infty_0 \rho_{NFW}\ dl \nonumber \\
&=&  \frac{2r_s\rho_s}{x^2-1}(1-\xi) ,
\end{eqnarray} 
\begin{eqnarray}
\overline{\Sigma}(<x) &=& \frac{2}{x^2} \int^x_0 x' \Sigma(x)\ dx' \nonumber \\
&=&  \frac{4r_s\rho_s}{x^2}\left[\xi+\ln\left(\frac{x}{2}\right)\right],
\end{eqnarray} 
where $x=R/r_s$, $R$ is the projected radial distance from the lens centre and $\xi$ is a 4th order power series expansion as $x$ approaches 1. 
\begin{equation}
\xi = \begin{cases} 
\frac{2}{\sqrt{1-x^2}}\textrm{arctanh} \sqrt{\frac{1-x}{x+1}} & \textrm{if } x<0.98, \\
\frac{2}{\sqrt{x^2-1}}\textrm{arctan} \sqrt{\frac{x-1}{x+1}} & \textrm{if } x>1.02, \\
1 - \frac{2}{3} (x-1) + \frac{7}{15} (x-1)^2 - \frac{12}{35} (x-1)^3 + \frac{166}{630} (x-1)^4 &\textrm{otherwise.}
\end{cases}
\end{equation}

\clearpage

\begin{figure}
\centerline{
\includegraphics[width=80mm]{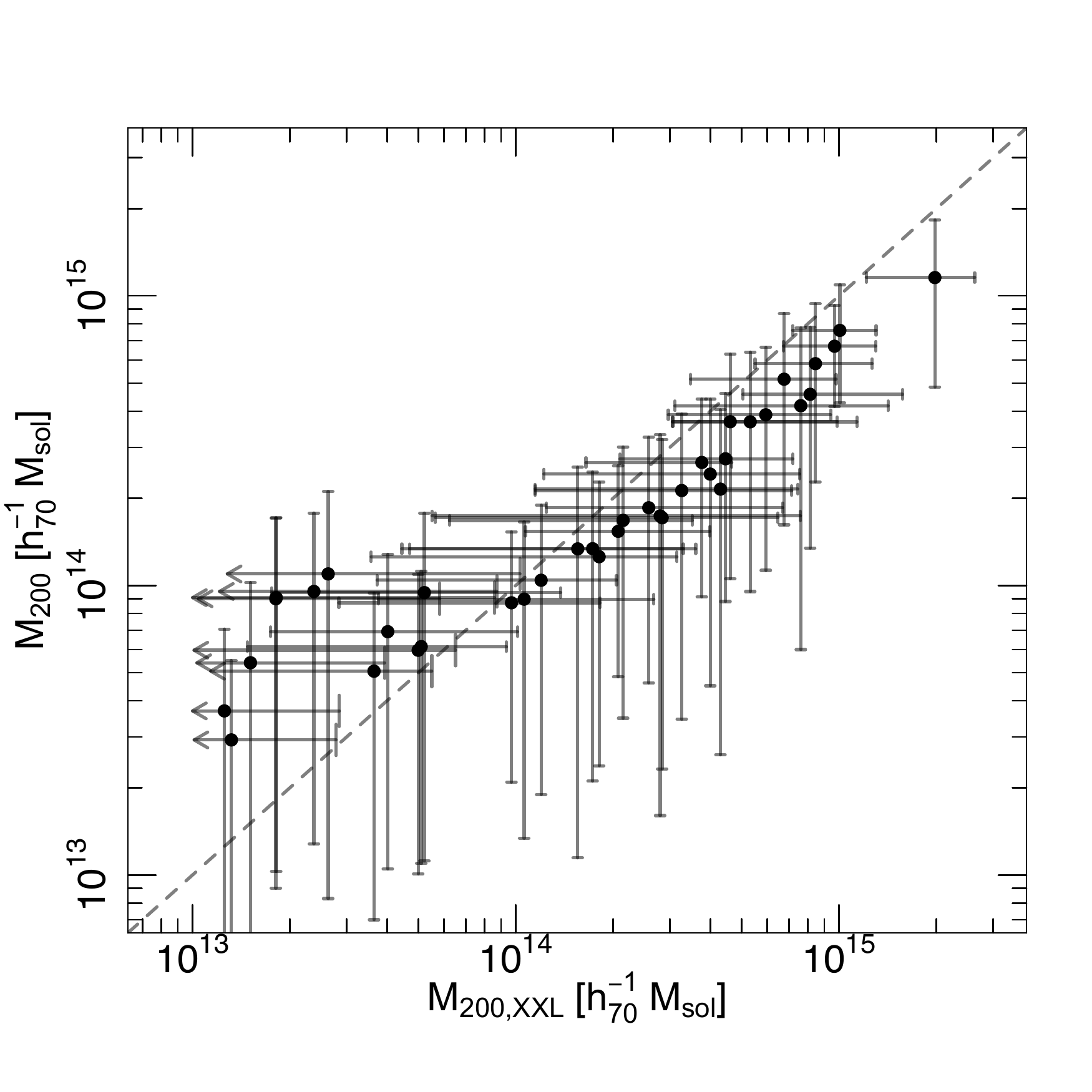}
}
\caption{Comparison between our masses and those measured within \protect\cite{Lieu2016} where they assume a fixed c--M relation from \protect\cite{Duffy2008}. Their upper limit measurements are shown in grey, where the estimate is confined by the lower prior boundary of 1$\times10^{13}\ \rm M_\odot$. The dashed line is equality. Our mass estimates show a systematic difference that is expected from the shrinking nature of the hierarchical model in that for high mass clusters we predict lower masses and low mass groups we predict higher mass values. The influence of the population distribution is more pronounced for the low mass systems where the uncertainties on the data are larger. \label{fig:plot_compare_old}}
\end{figure}

\begin{table}
\centering
\begin{tabular}{llll | llll}
\hline
\multicolumn{4}{l |}{Toy data} & \multicolumn{4}{l}{Observational data} \\
\hline
\hline
& prior  & & fit value & & prior  & & fit value \\
\hline
$\mu_0[1] =$ 	& 30 & $\overline{\mu_1}$ =	& 33.11$\pm$ 0.18 		& $\mu_0[1]$ = & 30 &  $\overline{\mu_1}$ = 	& 32.53 $\pm$0.33\\
		     	& 31 & 					& 33.14$\pm$ 0.18 		& 			& 31 &					& 32.64 $\pm$0.28\\
		     	& \textbf{32} & 				& \textbf{33.18$\pm$ 0.19} & 			& \textbf{32} & 				& \textbf{32.72 $\pm$0.28}\\
			& 33 	& 					& 33.21$\pm$ 0.18 		& 			& 33	& 					& 32.80 $\pm$0.26\\
			& 34 & 					& 33.24$\pm$ 0.18 		& 			& 34	& 					& 32.86 $\pm$0.25\\
$\mu_0[2] =$ 	& -1 	& $\overline{\mu_2}$ =	& 0.96$\pm$ 0.09 		& $\mu_0[2]$ = & -1 &  $\overline{\mu_1}$ = 	& 0.44 $\pm$ 0.42\\
			& 0	& 					& 0.97$\pm$ 0.09 		& 			& 0 	& 					& 0.57 $\pm$ 0.36\\
			& \textbf{1} & 				& \textbf{0.98$\pm$0.09} 	& 			& \textbf{1} & 				& \textbf{0.71$\pm$ 0.36}\\
			& 2 	& 					& 0.99$\pm$ 0.09 		& 			& 2 & 					& 0.79 $\pm$ 0.34\\
\end{tabular}
\caption{The results of the tests on the assumed priors. The table shows fitted values of the population means for various prior central values. \label{tab:testpriors}}
\end{table}

\begin{figure}
\includegraphics[width=180mm]{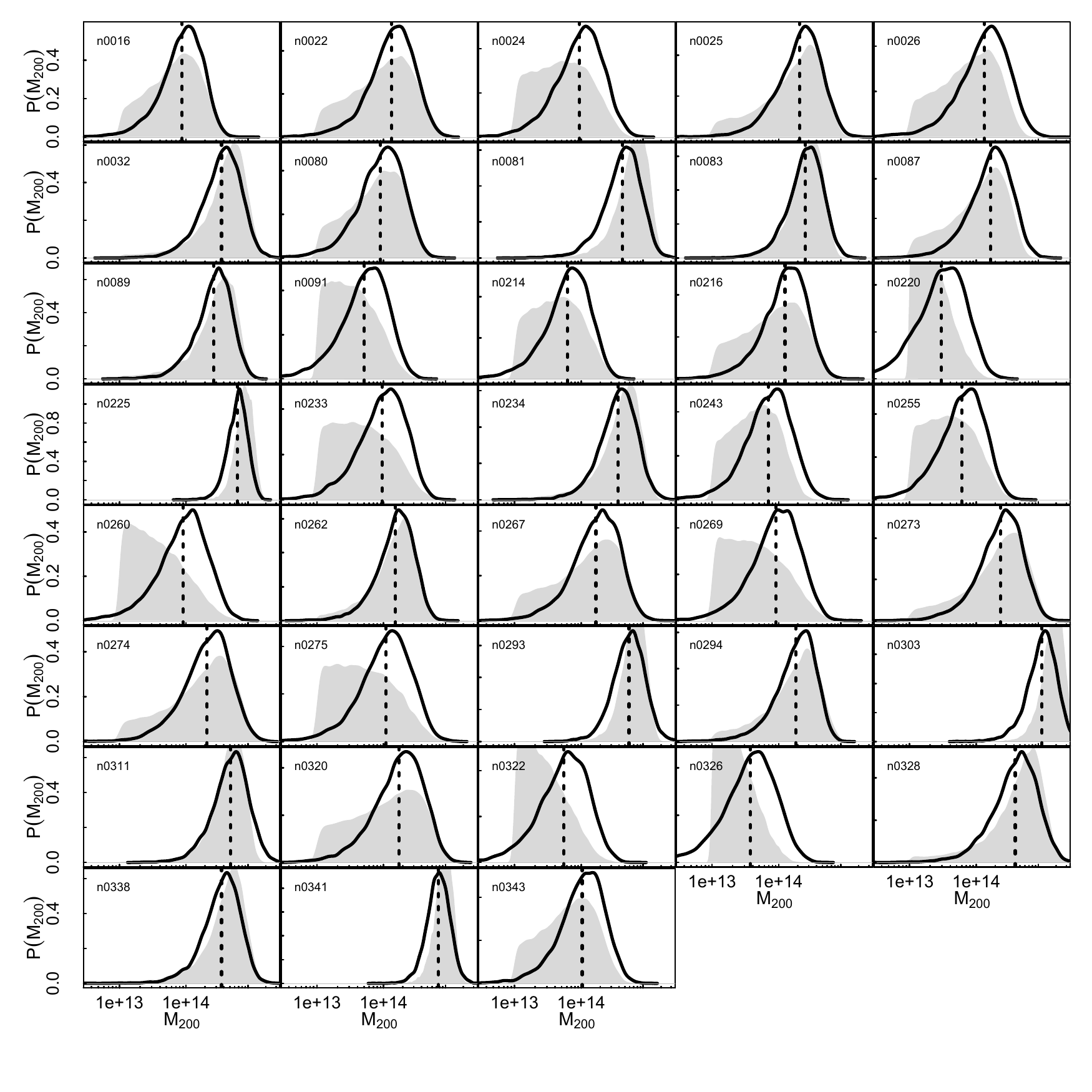}
 \caption{The posterior distribution functions of the individual mass measurements (solid black line) and the fit statistic taken as the posterior mean (dotted black line). The grey shaded regions show the posteriors of the individual masses from \protect\cite{Lieu2016} assuming a free concentration parameter for comparison. The posteriors of both methods are in reasonable agreement. The truncated prior used in \protect\cite{Lieu2016} can be seen at $10^{13}\ \rm M_\odot$ for their clusters where only an upper limit on mass is measured, whereas our posteriors do not incur a sharp prior boundary yet are still able to constrain a posterior peak.  
 \label{fig:plot_compare_old_mass_posteriors}}
\end{figure}

\end{document}